\documentclass[letterpaper,12pt]{article} 
\usepackage{amsmath}
\usepackage{amssymb}
\usepackage[round]{natbib}
\usepackage[dvips]{epsfig}
\usepackage{dcolumn}
\usepackage{enumerate}
\usepackage{hhline}
\usepackage{dsfont}
\usepackage{afterpage}
\usepackage{arydshln}
\usepackage{graphicx}
\usepackage{color}
\usepackage[usenames,dvipsnames]{xcolor}
\usepackage{rotating}
\usepackage[breaklinks]{hyperref}
\usepackage{xr}
\usepackage[percent]{overpic}
\usepackage{subfig}
\usepackage[]{caption}
\usepackage[ruled,vlined]{algorithm2e}
\usepackage{diagbox}
\usepackage{graphicx}
\usepackage{wrapfig}
\usepackage{lscape}
\usepackage{mathabx}
\usepackage{adjustbox}
\input 
\usepackage{fontenc}
\usepackage{setspace}
\usepackage{bm}
\usepackage{slashbox}
\usepackage{lscape}
\usepackage{multirow}
\usepackage{eurosym}
\usepackage{titlesec}
\usepackage{placeins}
\usepackage[mdyy]{datetime}
\epsfverbosetrue
\def\theequation{\thesection.\arabic{equation}}  
\def\abstract{\if@twocolumn
\section*{Abstract}
\else \normalsize 
\begin{center}
{\bf Summary\vspace{-.5em}\vspace{0pt}} 
\end{center}
\quotation 
\fi}
\def\endabstract{\if@twocolumn\else\endquotation\fi}

\makeatletter
\newcommand{\myappendix}[1]{
	\setcounter{section}{1}
        \renewcommand{\thesection}{A\arabic{section}}}

\DeclareMathAlphabet{\mathpzc}{OT1}{pzc}{m}{it}
\DeclareFontFamily{OT1}{pzc}{}
\DeclareFontShape{OT1}{pzc}{m}{it}{<-> s * [1.200] pzcmi7t}{}
\def \dsE {\text{$\mathds{E}$}}
\def \dsR {\text{$\mathds{R}$}}

\DeclareMathOperator{\diag}{diag}

\DeclareMathOperator*{\argmax}{{arg\,max}}

\DeclareMathOperator{\CD}{\mathcal{C}}
\DeclareMathOperator{\HCD}{\mathcal{HC}}
\DeclareMathOperator{\ND}{\mathcal{N}}

\DeclareMathOperator{\UD}{\mathcal{U}}

\DeclareMathOperator{\GaD}{\mathcal{G}}

\DeclareMathOperator{\IGD}{\mathcal{IG}}

 \newcommand{\Dir}{\mathpzc{Dir}}
  
\def\diag{{\mbox{diag}}}
\def \dsR {\text{$\mathds{R}$}}

\def\pkl{p_k^{(l)}}
\def\Kl{K^{(l)}}
\def\zil{z_i^{(l)}}
\def\zilm1{z_i^{(l-1)}}
\def\mukl{\mu_k^{(l)}}
\def\Bkl{B_k^{(l)}}
\def\Bl{B^{(l)}}
\def\epsikl{\epsilon_{ik}^{(l)}}
\def\delkl{\Delta_k^{(l)}}
\def\Dl{D^{(l)}}
\def\Dlm1{D^{(l-1)}}
\def\delkl{\delta_k^{(l)}}
\def\zl{z^{(l)}}
\def\mul{\mu^{(l)}}
\def\vec{\text{vec}}
\def\kapl{\kappa^{(l)}}
\usepackage{color}
\usepackage{colordvi}
\newlength{\breite}
\breite\textwidth
\newcounter{aufg}[section]
  {\refstepcounter{aufg}\noindent\textbf{Exercise \arabic{aufg}:}
   \\*[1ex]\noindent}{\vspace{.5cm}}
   
 \newcounter{notes}[section]
  {\refstepcounter{aufg}\noindent\textbf{}
   \\*[1ex]\noindent}{\vspace{.5cm}}
   
\usepackage{amsthm}

\makeatother

\newtheoremstyle{break}% name
  {}%         Space above, empty = `usual value'
  {}%         Space below
  {}% Body font
  {}%         Indent amount (empty = no indent, \parindent = para indent)
  {\bfseries}% Thm head font
  {.}%        Punctuation after thm head
  {\newline}% Space after thm head: \newline = linebreak
  {}%         Thm head spec
  
\theoremstyle{break}

\usepackage[margin=1in]{geometry}
\titlespacing*\section{0pt}{0pt plus 4pt minus 2pt}{0pt plus 2pt minus 2pt}
\titlespacing*\subsection{0pt}{0pt plus 4pt minus 2pt}{0pt plus 2pt minus 2pt}
\titlespacing*\subsubsection{0pt}{0pt plus 4pt minus 2pt}{0pt plus 2pt minus 2pt}

\newcounter{myremark}

\newcounter{mynotation}

\makeatletter
\def\@seccntformat#1{\@ifundefined{#1@cntformat}%
	{\csname the#1\endcsname\quad}  % default
	{\csname #1@cntformat\endcsname}% enable individual control
}

\usepackage{titlesec}

\usepackage{scalerel,stackengine}
\stackMath
\newcommand\reallywidehat[1]{%
\savestack{\tmpbox}{\stretchto{%
  \scaleto{%
    \scalerel*[\widthof{\ensuremath{#1}}]{\kern-.6pt\bigwedge\kern-.6pt}%
    {\rule[-\textheight/2]{1ex}{\textheight}}%WIDTH-LIMITED BIG WEDGE
  }{\textheight}% 
}{0.5ex}}%
\stackon[1pt]{#1}{\tmpbox}%
}

\begin{document}
\setlength{\abovedisplayskip}{0.15cm}
\setlength{\belowdisplayskip}{0.15cm}
\pagestyle{empty}
%\singlespacing
\begin{titlepage}

\title{
 Variational Inference and Sparsity in High-Dimensional Deep Gaussian Mixture Models
}

\author{Lucas Kock, Nadja Klein$^\ast$ and David J. Nott\\
%\small $^1$Humboldt  Universit\"at zu Berlin, $^2$National University of Singapore\\
%$^3$Melbourne Business School, University of Melbourne
}
\date{}
\maketitle

\vfill
\noindent
{\small Lucas Kock is PhD Student at Humboldt Universit\"at zu Berlin. Nadja Klein is Assistant Professor and  Emmy Noether Research Group Leader in Statistics and Data Science at Humboldt-Universit\"at zu Berlin.  David J. Nott is Associate Professor of Statistics and Applied Probability at National University of Singapore.\\
$^\ast$ Correspondence should be directed to~Prof.~Dr.~Nadja Klein at Humboldt Universit\"at zu Berlin, Emmy Noether Research Group in Statistics and Data Science,
Unter den Linden 6, 10099 Berlin, Germany. Email: nadja.klein@hu-berlin.de.}
%\textbf{code availability for review?!}

\vspace{2em}

\noindent {\small\textbf{Acknowledgments:} 
This work was supported by an NUS/BER Research Partnership grant by the National University of Singapore and Berlin University Alliance. Nadja Klein received support  by the German research foundation (DFG) through the  Emmy Noether grant KL 3037/1-1. 
}

%\normalsize
\newpage
\begin{center}
\mbox{}\vspace{2cm}\\
{\LARGE \title{
%Deep Gaussian Mixture Models with Variational Inference and Sparsity
%High-dimensional Clustering based on Variational Inference and Sparsity Priors for Deep Gaussian Mixture Models
Variational Inference and Sparsity in High-Dimensional Deep Gaussian Mixture Models
}
}\\
\vspace{1cm}
{\Large Abstract}
\end{center}
\vspace{-1pt}
\onehalfspacing
\noindent Gaussian mixture models are a popular tool for model-based clustering, and mixtures of factor analyzers are Gaussian mixture models having parsimonious factor covariance structure for mixture components.  There are several recent extensions of mixture of factor analyzers to deep mixtures, where the Gaussian model for the latent factors is replaced by a mixture of factor analyzers.  This construction can be iterated to obtain a model with many layers.  These deep models are challenging to fit, and we consider Bayesian inference using sparsity priors to further regularize the estimation.  A scalable natural gradient variational inference algorithm is developed for fitting the model, and we suggest computationally efficient approaches to the architecture choice using overfitted mixtures where unnecessary components drop out in the estimation.  In a number of simulated and two real examples, we demonstrate the versatility of our approach for high-dimensional problems, and demonstrate that the use of sparsity inducing priors can be helpful for obtaining improved clustering results.    
   
\vspace{20pt}
 
\noindent
{\bf Keywords}: Deep clustering; high-dimensional clustering; horseshoe prior; mixtures of factor analyzers; natural gradient; variational approximation.
\end{titlepage}

%\doublespacing

\newpage
\pagestyle{plain}
\setcounter{equation}{0}
\renewcommand{\theequation}{\arabic{equation}}
% !TeX spellcheck = en_US

\section{Introduction}\label{sec:intro}
Exploratory data analysis tools such as cluster analysis need to be increasingly flexible to deal with the greater complexity of datasets arising in modern applications.  For high-dimensional data, there has been much recent interest in deep clustering methods based on latent variable models with multiple layers.  Our work considers deep Gaussian mixture models for clustering, and in particular a deep mixture of factor analyzers (DMFA) model  recently introduced by \cite{viroli2019deep}.  The DMFA model is highly parametrized, and to make the estimation tractable \cite{viroli2019deep} consider a variety of constraints on model parameters which enable them to obtain promising results.  The objective of the current work is to consider a Bayesian approach to estimation of the DMFA model which uses sparsity-inducing priors to further regularize the estimation.  We use the horseshoe prior of \cite{CarPol2010}, and demonstrate in a number of problems that the use of sparsity-inducing priors is helpful. 
This is particularly true in clustering problems which are high-dimensional and involve a large number of noise features.  A computationally efficient natural gradient variational inference
scheme is developed, which is scalable to high-dimensions and large datasets.  
A difficult problem in the application of the DMFA model is the choice of architecture, i.e.~determining the number of layers and the dimension of each layer, since trying 
many different architectures is computationally burdensome.  We discuss useful heuristics for choosing good models and selecting the number of clusters in a computationally thrifty way, using overfitted mixtures \citep{rousseau2011asymptotic} with suitable priors on the mixing weights.

There are several suggested methods in the literature for constructing deep versions of Gaussian mixture models (GMM)s involving multiple layers of latent variables.  One of the earliest is the mixture of mixtures model of \cite{Li2005}.  \cite{Li2005} considers modelling non-Gaussian components in mixture models using a mixture of Gaussians, resulting in a mixture of mixtures structure.  The model is not identifiable through the likelihood, and \cite{Li2005} suggests several ways to address this issue, as well as methods for choosing the number of components for each cluster.  \cite{Malsiner-Walli+fg2017} identify the model through the prior in a Bayesian approach, and consider Bayesian methods for model choice.  Another deep mixture architecture is suggested in \cite{Ord+s2014}.  The authors consider a network with linear transformations at different layers, and the random sampling of a path through the layers.   Each path defines a mixture component by applying the corresponding sequence of transformations to a standard normal random vector.  Their approach allows a large number of mixture components, but without an explosion of the number of parameters due to the way  parameters are shared between components. 

The model of \cite{viroli2019deep} considered here is a multi-layered extension of the mixture of factor analyzers (MFA) model \citep{Ghahramani+h1997,mclachlan2003modelling}.  The MFA model is a GMM in which component covariance matrices have a parsimonious factor structure, making this model suitable for high-dimensional data.  The factor covariance structure can be interpreted as explaining dependence in terms of a low-dimensional latent Gaussian variable. 
Extending the MFA model to a deep model with multiple layers has two main advantages. First, non-Gaussian clusters can be obtained and second, it enables the fitting of GMMs with a large number of components, which is particularly relevant for high-dimensional data and cases when the number of true clusters is large.
\cite{Tang+sh2012} were the first to consider a deep MFA model, where a mixture of factor analyzers model is used as a prior for the latent factors instead of a Gaussian distribution.  Applying this idea recursively leads to deep models with many layers.  Their architecture splits components into  several subcomponents at the next layer in a tree-type structure. \citet{yang+hz17} consider a two layer version of the model of \cite{Tang+sh2012}  incorporating a common factor loading matrix at the first level and some other restrictions on the parametrization.  

The model of \cite{viroli2019deep} combines some elements of the models of \cite{Tang+sh2012}  and \cite{Ord+s2014}.   Similar to \cite{Tang+sh2012} an MFA prior is considered for the factors in an MFA model and multiple layers can be stacked together.  However, similar to the model of \cite{Ord+s2014}, their architecture has parameters for Gaussian component distributions with factor structure arranged in a network, with each Gaussian mixture component corresponding to a path through the network.  Parameter sharing between components and the factor structure makes the model parsimonious.  \cite{viroli2019deep} consider restrictions on the dimensionality of the factors at different layers and other model parameters to help identify the model.  A stochastic EM algorithm is used for estimation, and they report improved performance compared to greedy layerwise fitting algorithms.   For choosing the network architecture, they fit a large number of different architectures and use the Akaike Information Criterion (AIC) or Bayesian Information Criterion (BIC) to choose the final model.  \cite{selosse2020bumpy} report that even with the identification restrictions suggested in \cite{viroli2019deep} it can be very challenging to fit the DMFA model.  The likelihood surface has a large number of local modes, and it can be hard to find good modes, possibly because the estimation of a large number of latent variables is difficult.  This is one motivation
for the sparsity priors we introduce here.  While sparse Bayesian approaches have been considered before for both
factor models (\cite{carvalho2008,rockova+g16,hahn+hl18} among many others) and the MFA model \citep{ghahramani+b00}, 
they have not been considered for the DMFA model.

The structure of the paper is as follows. In Section~\ref{sec:MERGM}  we introduce the DMFA model of \cite{viroli2019deep} and our Bayesian treatment making appropriate prior choices.  Section~\ref{sec:VB} describes our scalable natural gradient variational inference method, and the approach to choice of the model architecture.  Section~\ref{sec:Simul} illustrates the good performance of our method for simulated data in a number of scenarios 
and compares our method to several alternative methods from the literature for simulated and benchmark datasets considered in \cite{viroli2019deep}.  
Although it remains challenging to fit complex latent variable models such as the DMFA, we find that in many situations which are well-suited to sparsity our approach is helpful for producing better clusterings. 
Sections~\ref{sec:gene} and \ref{sec:taxi} consider two real high-dimensional examples on  gene expression data and taxi networks illustrating the potential of our Bayesian treatment of DMFA models.  Section~\ref{sec:conclusion} gives some concluding discussion.   
\FloatBarrier
\section{Bayesian Deep Mixture of Factor Analyzers}\label{sec:MERGM}
This section describes our Bayesian DMFA model.  We consider a conventional MFA model first in Section \ref{sec:mfa}, and then the deep extension of this by \cite{viroli2019deep} in Section \ref{subsec:dgmm}. The priors used for Bayesian inference are discussed in Section \ref{subsec:priors}.

\subsection{MFA model}\label{sec:mfa}
In what follows we write $\ND(\mu,\Sigma)$ for the normal distribution with mean vector $\mu$ and 
covariance matrix $\Sigma$, and $\phi(y;\mu,\Sigma)$ for the corresponding density function.  
Suppose $y_i=(y_{i1},\dots, y_{id})^\top$, $i=1,\dots, n$ are independent identically distributed observations of dimension $d$.  
An MFA model for $y_i$ assumes a density of the form
\begin{align}
      & \sum_{k=1}^K p_k \phi(y_i;\mu_k,B_kB_k^\top+\delta_k), \label{mfa}
\end{align}
where $p_k>0$ are mixing weights, $k=1,\dots, K$, with $\sum_{k=1}^K p_k=1$; $\mu_k=(\mu_{k1},\dots, \mu_{kd})^\top$ are component
mean vectors; and $B_kB_k^\top+\delta_k$, $k=1,\dots, K$, are component specific covariance matrices with $B_k$  $d\times D$ matrices, $D<<d$, and 
$\delta_k$ are $d\times d$ diagonal matrices with diagonal entries $(\delta_{k1},\dots, \delta_{kd})^\top$.  The model \eqref{mfa} 
has a generative representation in terms of $D$-dimensional latent Gaussian random vectors, $z_i\sim \ND(0,I)$, $i=1,\dots, n$. 
Suppose that $y_i$ is generated with probability $p_k$ as
\begin{align}
y_i & =\mu_k+B_k z_i+\epsilon_{ik}, \label{mfa-generative}
\end{align}
where $\epsilon_{ik}\sim \ND(0,\delta_k)$.  Under this generative model, $y_i$ has density \eqref{mfa}.  The latent variables $z_i$ are 
called factors, and the matrices $B_k$ are called factor loadings or factor loading matrices.  In ordinary factor analysis \citep{Bartholomew2011}, which corresponds to $K=1$, restrictions on the factor loading matrices are made to ensure identifiability, and similar restrictions are needed
in the MFA model. Common restrictions are lower-triangular structure with positive diagonal elements, 
or orthogonality of the columns of the factor loading matrix. With continuous shrinkage sparsity-inducing priors and careful initialization of
the variational optimization algorithm, we did not find imposing an identifiability restriction such as a lower-triangular structure on the factor loading matrices to be necessary in our applications later.
Identifiability issues for sparse factor models with point mass mixture priors have been considered 
recently in \cite{FruLop2018}.

The MFA model is well-suited to analyzing high-dimensional data, because the modelling of dependence in terms of low-dimensional
latent factors results in a parsimonious model.  The idea of deep versions of the MFA model is to replace the Gaussian assumption $z_i\sim \ND(0,I)$ with the assumption that the $z_i$'s themselves follow an MFA model.  This idea can be applied recursively, to define a deep
model with many layers.

\subsection{DMFA model}\label{subsec:dgmm}
Suppose once again that 
$y_i$, $i=1,\dots, n$, are independent and identically distributed observations of dimension $D^{(0)}=d$, and 
define $y_i=z_i^{(0)}$.  We consider a hierarchical model in which latent 
variables $\zil$, at layer $l=1,\dots, L$ are generated according to the following generative process. Let $\Kl$ denote the number of mixture components at layer $l$. Then, with
probability $\pkl$, $k=1,\dots, \Kl$, $\sum_k \pkl=1$,  $	\zilm1$ is generated as 
\begin{align}
\zilm1 & = \mukl+\Bkl\zil+\epsikl,  \label{factormodel}
\end{align}
 where $\epsikl\sim \ND(0,\delkl)$, $\mukl$ is a $\Dlm1$-vector, $\Bkl$ is a $\Dlm1\times \Dl$ factor loading matrix, 
$\delkl=\diag(\delta_{k1}^{(l)},\dots, \delta_{k\Dlm1}^{(l)})$ is a $\Dlm1\times \Dlm1$ diagonal matrix with diagonal elements $\delta_{kj}^{(l)}>0$ and $z_i^{(L)}\sim \ND(0,I_{D^{(L)}})$.  Eq.~\eqref{factormodel} is the same as the generative model
\eqref{mfa-generative} for $\zilm1$, except that we have replaced the Gaussian assumption for the factors appearing 
on the right-hand side with a recursive modelling
using the MFA model.  Write $\vec(B_k^{(l)})$ for the vectorization of $B_k^{(l)}$, the vector obtained by stacking the elements
of $B_k^{(l)}$ into a vector columnwise.  Write
\begin{equation*}
\begin{aligned}[c]
\zl & = ({z_1^{(l)}}^\top, \dots, {z_n^{(l)}}^\top)^\top, \\
\mul & = ({\mu_1^{(l)}}^\top,\dots, {\mu_{\Kl}^{(l)}}^\top)^\top, \\
\Bl & =({\vec (B_1^{(l)})}^\top,\dots, {\vec(B_{\Kl}^{(l)})}^\top)^\top, \\
\delta^{(l)} & =({\delta_1^{(l)}}^\top,\dots, {\delta_{\Kl}^{(l)}}^\top)^\top \\
p^{(l)} & =(p_1^{(l)},\dots, p_{\Kl}^{(l)})^\top,
\end{aligned} \qquad
\begin{aligned}[c]
z & = ({z^{(1)}}^\top,\dots, {z^{(L)}}^\top)^\top, \\
\mu & =({\mu^{(1)}}^\top,\dots, {\mu^{(L)}}^\top)^\top, \\
B & =({B^{(1)}}^\top,\dots, {B^{(L)}}^\top)^\top, \\
\delta & =({\delta^{(1)}}^\top,\dots, {\delta^{(L)}}^\top)^\top, \\
p & =({p^{(1)}}^\top,\dots, {p^{(L)}}^\top)^\top.
\end{aligned}
\end{equation*}
In  \eqref{factormodel}, we  also denote $\gamma_{ik}^{(l)}=1$ if $z_i^{(l-1)}$ is generated from the $k$-th
component model with probability $p_k^{(l)}$,  $\gamma_{ik}^{(l)}=0$ otherwise, and 
\begin{align*}
& \gamma=({\gamma^{(1)}}^\top,\dots, {\gamma^{(L)}}^\top)^\top, \;\;\; \gamma^{(l)}=({\gamma_1^{(l)}}^\top,\dots, {\gamma_n^{(l)}}^\top)^\top,\;\;\;
\gamma_i^{(l)}=(\gamma_{i1}^{(l)},\dots, \gamma_{i\Kl}^{(l)})^\top.
\end{align*}

\cite{viroli2019deep} observe that their DMFA model is just a Gaussian mixture model with $\prod_{l=1}^L K^{(l)}$ components.  The components correspond to ``paths" through
the factor mixture components at the different levels.  Write $k_l\in \{1,\dots, K^{(l)}\}$ for the index
of a factor mixture component at level $l$.  Let $k=(k_1,\dots, k_L)^\top$ index a path.  
Let $p(k)=\prod_{l=1}^L p_{k_l}^{(l)}$,
$$\mu(k)=\mu_{k_1}^{(1)}+\sum_{l=2}^L\left(\prod_{m=1}^{l-1}B_{k_m}^{(m)}\right)\mu_{k_l}^{(l)}
\;\;\mbox{ and }\;\;
\Sigma(k)=\delta_{k_1}^{(1)}+\sum_{l=2}^L \left(\prod_{m=1}^{l-1} B_{k_m}^{(m)}\right) \delta_{k_l}^{(l)}\left(\prod_{m=1}^{l-1}B_{k_m}^{(m)}\right)^\top.$$
Then the DMFA model corresponds to the Gaussian mixture density
$$\sum_{k} p(k) \phi(y;\mu(k),\Sigma(k)).$$
The  DMFA  allows expressive modelling through a large number of mixture components, but due to 
the parameter sharing and the factor structure the model remains parsimonious.

\subsection{Prior distributions}\label{subsec:priors}	

For our priors on component mean parameters at each layer, we use heavy-tailed Cauchy priors, and for
standard deviations, half-Cauchy priors.  These thick-tailed priors tend to be dominated by likelihood information
in the event of any conflict with the data.  For component factor loading matrices, we use sparsity-inducing
horseshoe priors.  Sparse structure in factor loading matrices is motivated by the common situation where dependence can be explained through latent factors, where each one influences only a small subset of components. 
If the dependence structure is complex, it may be helpful to consider non-sparse factor loadings at the first layer, with higher levels of shrinkage in the prior at deeper layers where an 
elaborate modelling of dependence structure for the latent variables may be difficult to sustain.  

\cite{Malsiner-Walli+fg2017} consider a Bayesian
approach to the mixture of mixtures model, which is a kind of DGMM, 
and give an interesting discussion of prior choice in that context.  
They observe that in the mixture of mixtures, the mixture at the higher level groups together the lower level 
clusters to accommodate non-Gaussianity, and their prior choices are constructed with this in mind.  
Unfortunately, the intuition behind their priors does not hold for the model considered here, since the 
network architecture of \cite{viroli2019deep} does not possess the nested structure of the mixture of mixtures,
which is crucial to the prior construction in \cite{Malsiner-Walli+fg2017}.  

A precise description of the priors will be given now.
Denote the the complete vector of parameters by $\vartheta  = (\mu^\top,B^\top,z^\top,p^\top,\delta^\top,\gamma^\top)^\top.$ 
 First, we assume independence between $\mu$, $B$, $z$, $\delta$ and $p$
in their prior distribution. For the marginal prior distribution for $\mu$, we furthermore assume independence between all components, and the marginal prior density for $\mu_{kj}^{(l)}$ is assumed to be a Cauchy density, $\CD(0,G^{(l)})$, where $G^{(l)}$ is a known hyperparameter possibly depending on $l$, $l=1,\dots, L$.  
This Cauchy prior can be represented hierarchically as a mixture of a univariate Gaussian and inverse gamma distribution, i.e.
\begin{align*}
& \mu_{kj}^{(l)}|g_{kj}^{(l)}\sim \ND(0,G^{(l)}g_{kj}^{(l)}),\;\;\;g_{kj}^{(l)}\sim \IGD\left(\frac{1}{2},\frac{1}{2}\right),
\end{align*}
and we write 
\begin{align*}
& g=({g^{(1)}}^\top,\dots, {g^{(L)}}^\top)^\top,\;\;\;g^{(l)}=({g_1^{(l)}}^\top,\dots, {g_{\Kl}^{(l)}}^\top)^\top,
& g_k^{(l)}=(g_{k1}^{(l)},\dots, g_{k\Dlm1}^{(l)})^\top.
\end{align*}
We  augment the  model to include the parameters $g$ in this hierarchical representation.  

For the prior on $B$, we assume elements are independently distributed {\it a priori} with a horseshoe prior~\citep{CarPol2010}, 
\begin{align*}
&\vec (\Bkl)_j \sim \ND\left(0,\tau_k^{(l)}/h_{kj}^{(l)}\right),\;  \tau_{k}^{(l)}\sim \IGD\left(\frac{1}{2},\frac{1}{\xi_k^{(l)}}\right),\; \xi_k^{(l)}\sim \IGD\left(\frac{1}{2},\frac{1}{(\nu^{(l)})^2}\right),\\
&h_{kj}^{(l)}\sim \GaD\left(\frac{1}{2},c_{kj}^{(l)}\right),\;c_{kj}^{(l)}\sim  \GaD\left(\frac{1}{2},1\right),
\end{align*}
where $\GaD(a,b)$ denotes a gamma distribution with shape $a$ and scale parameter $b$ and we write
\begin{equation*}
\begin{aligned}[c]
h & =({h^{(1)}}^\top,\dots, {h^{(L)}}^\top)^\top, \\
c & =({c^{(1)}}^\top,\dots, {c^{(L)}}^\top)^\top, 
\end{aligned}
\qquad
\begin{aligned}[c]
h^{(l)}& =({h_1^{(l)}}^\top,\dots, {h_{\Kl}^{(l)}}^\top)^\top, \\
c^{(l)} & =({c_1^{(l)}}^\top,\dots, {c_{\Kl}^{(l)}}^\top)^\top,
\end{aligned}
\qquad
\begin{aligned}[c]
h_k^{(l)} & =(h_{k1}^{(l)},\dots, h_{k\kapl}^{(l)})^\top, \\
c_k^{(l)} & =(c_{k1}^{(l)},\dots, c_{k\kapl}^{(l)})^\top,
\end{aligned}
\end{equation*}
\begin{equation*}
\begin{aligned}[c]
\tau & =({\tau^{(1)}}^\top,\dots, {\tau^{(L)}}^\top)^\top, \\
\xi & =({\xi^{(1)}}^\top,\dots, {\xi^{(L)}}^\top)^\top,
\end{aligned}
\qquad
\begin{aligned}[c]
\tau^{(l)} & =({\tau_1^{(l)}}^\top,\dots, {\tau_{\Kl}^{(l)}}^\top)^\top, \\
\xi^{(l)}& =({\xi_1^{(l)}}^\top,\dots, {\xi_{\Kl}^{(l)}}^\top)^\top,
\end{aligned}
\end{equation*}
where $\kapl=\Dlm1 \Dl$ and $\nu=({\nu^{(1)}}^\top,\dots, {\nu^{(L)}}^\top)^\top\in\dsR^{L}$ are scale parameters assumed to be known.  We  augment the  model to include the parameters $h,c,\tau,\xi$ in the hierarchical
representation.  

For $\delta$, we assume all elements are independent in the prior with 
$\sqrt{\delta_{kj}^{(l)}}$ being half-Cauchy distributed, $\sqrt{\delta_{kj}^{(l)}}\sim \HCD(A^{(l)})$, where the scale parameter
$A^{(l)}$ is known and possibly depending on $l$, $l=1,\dots, L$.  This prior can also be expressed hierarchically, 
\begin{align*}
& \delta_{kj}^{(l)}|\psi_{kj}^{(l)}\sim \IGD\left(\frac{1}{2},\frac{1}{\psi_{kj}^{(l)}}\right)\;\;\; \psi_{kj}^{(l)}\sim \IGD\left(\frac{1}{2},\frac{1}{{A^{(l)}}^2}\right),
\end{align*}
and we write
\begin{align*}
& \psi=({\psi^{(1)}}^\top,\dots, {\psi^{(L)}}^\top)^\top,\;\;\;\psi^{(l)}=({\psi_1^{(l)}}^\top,\dots, {\psi_{\Kl}^{(l)}}^\top)^\top,
& \psi_k^{(l)}=(\psi_{k1}^{(l)},\dots, \psi_{k\Dlm1}^{(j)})^\top.
\end{align*}
Once again, we can augment the original model to include $\psi$ and we do so.

The prior on $p$ assumes independence between $p^{(l)}$, $l=1,\dots, L$, and the marginal prior for $p^{(l)}$ is 
Dirichlet, $p^{(l)}\sim \Dir(\rho_1^{(l)},\dots, \rho_{\Kl}^{(l)})^\top$.  
 The full set of unknowns in the model is $\lbrace \mu, B, z, p, \delta,\gamma,p,\psi,g,h,c,\tau,\xi\rbrace$ and we denote the  corresponding vector of unknowns by 
\begin{align*}
\theta & = ( \mu^\top,\vec(B)^\top,z^\top,p^\top,\delta^\top,\gamma^\top,\psi^\top,g^\top,h^\top,c^\top,\tau^\top,\xi^\top)^\top.  
\end{align*}

\FloatBarrier
\section{Variational Inference for the Bayesian DMFA}\label{sec:VB}
Next we review basic ideas of variational inference and  introduce a scalable variational inference method for the DMFA model.
\subsection{Variational inference}

Variational inference (VI) approximates a posterior density $p(\theta|y)$ by assuming a form $q_\lambda(\theta)$ for it and then minimizing some measure of closeness to $p(\theta|y)$.  Here, $\lambda$ denotes variational parameters to be optimized, such as the mean
and covariance matrix parameters 
in a multivariate Gaussian approximation.  If the measure of closeness adopted is the Kullback-Leibler divergence, 
the optimal approximation maximizes the evidence lower bound (ELBO) 
\begin{equation}\label{eq:Elb}
\mathcal{L}(\lambda)=\int q_{\lambda}(\theta)\log\left\lbrace\frac{p(y,\theta)}{q_{\lambda}(\theta)}\right\rbrace d\theta\,,
\end{equation}
with respect to $\lambda$. Eq.~\eqref{eq:Elb} is  an expectation with respect to $q_\lambda(\theta)$,
\begin{equation}\label{eq:Elb2}
\mathcal{L}(\lambda)=\dsE_{q_\lambda}\left\lbrace\log(h(\theta))-\log(q_{\lambda}(\theta))\right\rbrace,\,
\end{equation}
where $h(\theta)=p(y|\theta)p(\theta)$ and this observation enables an unbiased Monte Carlo estimation of the gradient of $\mathcal{L}(\lambda)$ after differentiating under
the integral sign.  This is often used with stochastic gradient ascent methods to optimize the ELBO,  see \cite{Blei+km2017} for further background on VI methods.
 For some models and appropriate forms for $q_\lambda(\theta)$,  ${\cal L}(\lambda)$ can be expressed analytically, and 
this is the case for our proposed VI method for the DFMA model.  In this case, 
finding the optimal value $\lambda^*$ does not require 
Monte Carlo sampling from the approximation to implement the variational optimization 
\citep[see e.g.][]{honkela2007natural}, although the use of subsampling to deal with large
datasets may still require the use of stochastic optimization.

\subsection{Mean field variational approximation for DMFA}
Choosing the form of $q_\lambda(\theta)$ requires balancing flexibility and computational tractability.  Here we consider
the factorized form 
\begin{align} \label{eq:mfvb}
q_\lambda(\theta) & = q(\mu)q(\vec(B))q(z)q(p)q(\delta)q(\gamma)q(\psi)q(g)q(h)q(c)q(\tau)q(\xi).
\end{align}
and each factor in Eq.~\eqref{eq:mfvb} will also be fully factorized.  For the DMFA model, the existence of conditional conjugacy through our prior choices
means that the parametric form
of the factors follows from the factorization assumptions made. 
We can give an explicit expression for the ELBO (see  the Web Appendix {A.2} for details) and 
use numerical optimization techniques to find the optimal variational parameter $\lambda^*$.  It is possible to consider less restrictive factorization assumptions than we have, retaining some of the dependence structure, while at the same time retaining the closed form for the lower bound.  A fully factorized approximation has been used to reduce the number of variational parameters to optimize.  

\subsection{Approach to computations for large datasets} \label{sec: sgd}
Optimization of the lower bound \eqref{eq:Elb2} with respect to the high-dimensional vector of variational parameters
is difficult.  One problem is that the number of variational parameters grows with the sample size.  This occurs because we 
need to approximate posterior distributions of observation specific latent variables $z_i^{(l)}$ and 
$\gamma_i^{(l)}$, $i=1,\dots, n$, $l=1,\dots, L$.  To address this issue, we adapt the idea
of stochastic VI by \cite{hoffman2013stochastic}.  More specifically, we split  $\theta$ into so-called
``global'' and ``local'' parameters, $\theta=(\beta^\top,\zeta^\top)^\top$, where 
\begin{align*}
  \beta=(\mu^\top,B^\top,p^\top,\delta^\top,\psi^\top,g^\top,h^\top,c^\top,\tau^\top,\xi^\top)^\top,
\end{align*}
and $\zeta=(z^\top,\gamma^\top)^\top$ denote the ``global'' and ``local'' parameters of $\theta$, respectively.  

We can similarly partition the variational parameters $\lambda$ into global and local components, 
$\lambda=(\lambda_G^\top,\lambda_L^\top)^\top$, where $\lambda_G$ and $\lambda_L$ are the variational parameters
appearing in the variational posterior factors involving elements of $\beta$ and $\zeta$, respectively. 
Write the lower bound ${\cal L}(\lambda)$ as ${\cal L}(\lambda_G,\lambda_L)$.  
Write $M(\lambda_G)$ for the value of $\lambda_L$ optimizing
the lower bound with fixed global parameter $\lambda_G$.  When optimizing ${\cal L}(\lambda)$, the optimal value
of $\lambda_G$ maximizes
\begin{align*}
  \overline{\cal L}(\lambda_G) & = {\cal L}(\lambda_G,M(\lambda_G)).
\end{align*}
Next, differentiate $\overline{\cal L}(\lambda_G)$ to obtain
\begin{align}
  \nabla_{\lambda_G} \overline{\cal L}(\lambda_G) & = \nabla_{\lambda_G}{\cal L}(\lambda_G,M(\lambda_G))
    + \nabla_{\lambda_G}M(\lambda_G)^\top \nabla_M {\cal L}(\lambda_G,M(\lambda_G)) \nonumber \\
     & = \nabla_{\lambda_G}{\cal L}(\lambda_G,M(\lambda_G)). \label{derivative}
 \end{align}
The last line follows from $\nabla_M {\cal L}(\lambda_G,M(\lambda_G))=0$, since $M(\lambda_G)$ gives the optimal local variational
 parameter values for fixed $\lambda_G$.  Eq.~\eqref{derivative} says that the gradient 
 $\nabla_{\lambda_G} \overline{\cal L}(\lambda_G)$ can be computed by first optimizing to find $M(\lambda_G)$, and
 then computing the gradient of ${\cal L}(\lambda)$ with respect to $\lambda_G$ with $\lambda_L$ fixed at $M(\lambda_G)$.  
 
 We optimize $\overline{\cal L}(\lambda_G)$ using stochastic gradient ascent methods, which iteratively update 
 an initial value $\lambda_G^{(1)}$ for $\lambda_G$ for $t\geq 1$ by the iteration 
 $$\lambda_G^{(t+1)}=\lambda_G^{(t)}+a_t\circ \widehat{\nabla_{\lambda_G} \overline{\cal L}}(\lambda_G^{(t)}),$$
 where $a_t$ is a vector-valued step size sequence, $\circ$ denotes elementwise multiplication, and 
 $\widehat{\nabla_{\lambda_G} \overline{\cal L}}(\lambda_G^{(t)})$
 is an unbiased estimate of the gradient of $\nabla_{\lambda_G} \overline{\cal L}(\lambda_G)$ at $\lambda_G^{(t)}$.  
 An unbiased
 estimator of \eqref{derivative} is constructed by randomly sampling a mini-batch of the data.  This enables us to avoid
 dealing with all local variational parameters at once, lowering the dimension of any optimization.  
 Let $A$ be a subset of $\{1,\dots, n\}$ chosen uniformly at random without replacement inducing the set of indices
 of a data mini-batch.  
 Then, in view of \eqref{derivative}, an unbiased estimate $\widehat{\nabla_{\lambda_G}\overline{\cal L}}(\lambda_G)$ of 
 ${\nabla_{\lambda_G}}\overline{\cal L}(\lambda_G)$ can be obtained by 
  \begin{align*}
   \widehat{\nabla_{\lambda_G}\overline{\cal L}}(\lambda_G) & = \nabla_{\lambda_G}\mathcal{L}^F(\lambda_G)+\frac{n}{|A|}\sum_{i\in A} \nabla_{\lambda_G} \mathcal{L}^i(\lambda_G,M(\lambda_G)),
 \end{align*}
 where $|A|$ is the cardinality of $A$, and we have written ${\cal L}(\lambda)={\cal L}^F(\lambda_G)+\sum_{i=1}^n {\cal L}^i(\lambda)$, where ${\cal L}^i(\lambda)$ is the contribution to ${\cal L}(\lambda)$ of all terms involving the local parameter for the $i$th observation 
 and ${\cal L}^F(\lambda_G)$ is the remainder.
For this estimate, computation of all components of $M(\lambda_G)$ is not required, since only the optimal local variational
parameters for the mini-batch are needed.

To optimize $\lambda_G$ we use the natural gradient \citep{amari98} rather than the ordinary gradient.  
The natural gradient uses an update which respects the meaning of the variational parameters in terms of the
underlying distributions  they index.  
The natural gradient is given by $I_F(\lambda_G)^{-1} \widehat{\nabla_{\lambda_G}\overline{\cal L}}(\lambda_G)$, where  $I_F(\lambda_G)=\text{Cov}(\nabla_{\lambda_G} q_{\lambda_G}(\beta))$, where
$q_{\lambda_G}(\beta)$ is the $\beta$ marginal of $q_\lambda(\theta)$.
Because of the factorization of the variational posterior into independent components, it suffices to compute the submatrices of $I_F(\lambda_G)$ corresponding to each factor separately. 

\subsection{Algorithm}
The complete algorithm can be divided into two nested steps iterating over the local and global parameters, respectively. First, an update of the global parameters is considered using
a stochastic natural gradient ascent step for optimizing $\overline{\cal L}(\lambda_G)$, 
with the adaptive stepsize choice proposed in \cite{ranganath2013adaptive}.
For estimating the gradient in this step, the optimal local parameters $M(\lambda_G)$ for the current $\lambda_G$ need to be identified
for observations in a mini-batch. 
In theory, this can be done numerically by a gradient ascent algorithm. But this leads to a long run time, because one has to run this local gradient ascent until convergence for each step of the global stochastic gradient ascent. In our experience it is not necessary to calculate $M(\lambda_G)$ exactly, and it suffices to use an estimate, which helps to decrease the run time of the algorithm.
A natural approximate estimator for $M(\lambda_G)$ can be constructed in the following way.  Start by estimating the 
most likely path $\gamma_i$ using the clustering approach described in Section~\ref{sec: clustering}, while setting
$\beta$ to the current
variational posterior mean estimate for the global parameters.  The 
local variational parameters appearing in the factor $q(z_{ij}^{(l)})$ are then estimated layerwise starting with $l=1$ 
by setting this density proportionally to
$\exp(E(\log p(z_{ij}^{(l)}\vert z_{i}^{(l-1)},\gamma_i,\beta)))$, where the expectation is with respect to $q(z_{ij}^{(l-1)})$, 
and which has a closed form Gaussian expression.  
In layer 1, $z_i^{(0)}=y_i$.  The above approximation is motivated by the mean field update for $q(z_{ij})$, which is 
proportional to the exponential of the expected log full conditional for $z_{ij}^{(l)}$.  
However,  the usual mean field update to do the local optimization would
require iteration, which is why we use the fast layerwise approximation. The complete algorithm, details of the local optimization step and  explicit
expressions for the conditional $p(z_{ij}^{(l)}\vert z_i^{(l-1)},\gamma_i,\beta)$ are given in the Web Appendix {B.2}.
We calculate all gradients using the automatic differentiation of the Python package \texttt{PyTorch} \citep{paszke2017automatic}.

\subsection{Hyperparameter choices}
For all experiments we set the hyperparameters $G^{(l)}=2,\nu^{(l)}=1$ and $A^{(l)}=2.5$. The hyperparameter $\rho_k^{(l)}$ is set either to $1$ if the number of components in the respective layer is known, or to $0.5$ if the number of components is unknown, see Section~\ref{sec: model selection} for further discussion. The size of the mini-batch is set to $5\%$ of the data size, but not less then 1 or larger then 1024. This way the number of variational parameters which need to be updated at every step remains
bounded regardless of the sample size $n$.We found these hyperparameter choices to work well in all the diverse scenarios considered in Sections~\ref{sec:Simul}, \ref{sec:gene} and~\ref{sec:taxi}.

\subsection{Clustering with DMFA} \label{sec: clustering}
The algorithm described in the previous section returns optimal variational parameters $\lambda_G$ for the global parameters $\beta$. A canonical point estimator for $\beta$ is then given by the mode of $q_\lambda(\theta)$. 
Following \cite{viroli2019deep}, we consider only the first layer of the model for clustering. Specifically, the cluster of data point $y$ is given by 
$$\argmax_{k=1,\dots, K^{(1)}} p_k^{(1)} p(y\vert \beta,\gamma^{(1)}_k=1).$$  
By integration over the local parameters for $y$, $p(y \vert \beta,\gamma^{(1)}_k=1)$ can be written as a mixture of $\prod_{r=2}^L K^{(r)}$ Gaussians of dimension $D^{(0)}$ for $k=1,\dots, K^{(1)}$. This way, the parameters of the bottom layers can be viewed as defining a density approximation to the different clusters and the overall model can be interpreted as a mixture of Gaussian mixtures.
Alternatively, the DMFA can be viewed as a GMM with $\prod_{l=1}^LK^{(l)}$ components, where each component corresponds to a path through the model. By considering each of these Gaussian components as a separate cluster, the DMFA can be used to find a very large number of Gaussian clusters. However, \citet{selosse2020bumpy} note that some of these ``global'' Gaussian components might be empty, even when there are data points assigned to every component in each layer. We investigate the idea of using the DMFA to fit a large scale GMM further in Section~\ref{sec:taxi}. 

\subsection{Model architecture and model selection} \label{sec: model selection}
So far it has been assumed that the number of layers $L$, the number of components in each layer $K^{(l)}$, and the dimensions 
of each layer $D^{(l)}$, $l=1,\dots L$, are known. As this is usually not the case in practice, we now discuss  how to choose a suitable architecture.

If the number of mixture components $K^{(l)}$ in a layer is unknown, we initialize the model with a relatively large number of components, and set the Dirichlet prior hyperparameters on the component weights to 
$\rho_1^{(l)}=\dots=\rho_{K}^{(l)}=0.5$.  A large $K^{(l)}$ corresponds to an overfitted mixture, 
and for ordinary Gaussian mixtures unnecessary components will drop out under suitable conditions \citep{rousseau2011asymptotic}. 
A VI method for mixture models considering overfitted mixtures for model choice is discussed
in \cite{mcgrory2007variational}.  
The results of \cite{rousseau2011asymptotic} do not apply directly to deep mixtures because of the way parameters
are shared between components, but we find that this method for choosing components is useful in practice.
Our experiments show that in the case of deep Gaussian mixtures, the variational posterior concentrates around a small subset of components with high impact, while setting the weights for other components close to zero. Therefore, we remove all components with weights smaller than a threshold, set to be $0.01$ in  simulations. After  this, it is better 
in our experience to refit the model with the smaller number of components than to keep the fit initialized with the large number of components.

The choice of the number of layers $L$ and the dimensions $D$ is a classical model selection problem, where the model $m$ is chosen out of a finite set of  proposed models in the model space $\mathcal{M}$. We assume a uniform 
prior on the model space $p(m)\propto 1$ for each $m\in\mathcal{M}$. Hence, the model can be selected by 
$
    \hat{m}=\argmax_{m\in\mathcal{M}} p(m|y)=\argmax_{m\in\mathcal{M}} p(y|m).
$
Denoting the ELBO for the architecture $m$ with variational parameters $\lambda_m$ by 
$\mathcal{L}(\lambda_m \vert m)$, this is a lower bound for $\log p(y|m)$, which is tight if the variational posterior
approximation is exact.  Hence we choose the selected model by
%\begin{equation*}
    $\hat{m}=\argmax_{m\in\mathcal{M}} \mathcal{L}(\lambda_m^* \vert m),$
%\end{equation*}
where $\lambda_m^*$ denotes the optimal variational parameter value $\lambda_m$ for model $m$. Since it is computational expensive to run the VI algorithm until convergence for all models $m\in\mathcal{M}$, a naive approach would be to estimate $\mathcal{L}(\lambda_m^* \vert m)$ via $\mathcal{L}(\lambda_m^{(T)}\vert m)$ after running the algorithm for $T$ steps, 
where $\lambda_m^{(t)}$ is the value for $\lambda_m$ at iteration $t$.
In our experiments we estimated $\mathcal{L}(\lambda_m^*\vert m)$ as the mean over the last $5\%$ of $250$ iterations, i.e.
%\begin{equation*}
$    \hat{m}=\argmax_{m\in\mathcal{M}} \sum_{t=238}^{250} \mathcal{L}(\lambda_m^{(t)}\vert m).$
%\end{equation*}
This calculation can be be parallelized  by running the algorithm for each model $m$ on a different machine and only involves evaluations of the ELBO, which does not induce additional computational burden as the ELBO needs to be calculated for fitting the model. For the dimensions we test all possible choices fulfilling the Anderson-Rubin condition $D^{(l+1)}\leq \frac{D^{(l)}-1}{2}$ for $l=0,\dots,L$, which is a necessary condition for model identifiability \citep{FruLop2018}. Additionally, this condition gives an upper bound for the number of layers. However, in our experiments, we consider only models with $L=2$ or $L=3$ layers, because in our experience architectures with few layers and a rapid decrease in dimension outperform architectures with many deep layers. Once an architecture is chosen based on short runs, the fitting algorithm is
fully run to convergence for the optimal choice.
\FloatBarrier
\section{Benchmarking using simulated and real data}\label{sec:Simul}
To demonstrate the advantages of our Bayesian DMFA with respect to clustering high-dimensional non-Gaussian data, computational efficiency of model selection, accommodating sparse structure and scalability to large datasets, we experiment on several simulated and publicly available benchmark examples.  

\subsection{Design of numerical experiments}
First we consider two simulated datasets where 
our Bayesian approach with sparsity-inducing priors is able to outperform the maximum likelihood method considered in
\cite{viroli2019deep}.  The data generating process for the first  dataset scenario S1 is a GMM with many noisy (uninformative) features and sparse covariance matrices. Here, sparsity priors are helpful because there are many noise features.  The data in scenario S2 has a similar structure to the real data from the application presented in Section~\ref{sec:gene} having unbalanced non-Gaussian clusters. Here, the true
generative model has highly unbalanced group sizes, and the regularization that our priors provide is useful in this setting for
stabilizing the estimation. 

Five real datasets considered in \citet{viroli2019deep} are tested as well, and our method shows similar performance to theirs for these examples. However, we compare our method, which we will label VIdmfa, not only to the approach of  \citet{viroli2019deep} (EMdgmm) but to several other benchmark methods, including a GMM based on the EM algorithm (EMgmm) and Bayesian VI approach by \citet{mcgrory2007variational} (VIgmm), a skew-normal mixture (SNmm), a skew-t mixture (STmm), k-means (kmeans), partition around metroids (PAM), hierarchical clustering with Ward distance (Hclust), a factor mixture analyser (FMA) and a mixture of factor analyzer (MFA). To measure how close a set of clusters is to the ground truth labels, 
we consider three popular performance  measures (e.g.~\cite{vinh2010information}).  These are 
the missclassification rate (MR), the adjusted rand index (ARI) and the adjusted mutual information (AMI).% We refer to the Web Appendix {C} for full details on datasets, benchmark methods and measures of performance and focus on the main results in the following.

\subsection{Datasets}

Below we describe the two simulated scenarios S1 and S2 involving sparsity,  as well
as the five real datasets used in
\cite{viroli2019deep}.\\
\textbf{Scenario S1: Sparse location scale mixture.} Datasets with $D^{(0)}=30$ features are drawn from a mixture of high-dimensional Gaussian distributions with sparse covariance structure. The first $15$ of the features yield information on the clusters and are obtained from a mixture of $K=5$ Gaussian distributions with different means $\mu_k$ and covariances $\Sigma_k$. The mean for component $k=1,\ldots,5$ has entries
\begin{equation*}
(\mu_k)_j=
\begin{cases}                                   
-1 & \mbox{ if $k$ divides $j$} \\
1 & \mbox{ otherwise,} 
\end{cases} 
\end{equation*}
for $j=1,\dots,15$. The covariance matrices $\Sigma_k$ are drawn independently based on the Cholesky factors via the function \texttt{make\_sparse\_spd\_matrix} from the Python package \texttt{scikit-learn} \citep{scikit-learn}. An entry of the Cholesky factros of $\Sigma_k$ is set to zero  with a probability $\alpha=0.95$ and all non-zero entries are assumed to be in the interval $[0.4,0.7]$. %, so that non-zero components are not too close to zero.
The $15$ noise (irrelevant) features are drawn independently from an $\ND(0,1)$ distribution. The number of elements in each cluster is discrete uniformly distributed as $\UD\lbrace 50,200\rbrace$ leading to datasets of sizes $n\in[250,1000]$.
This clustering task is not as easy as it might first appear, since only half of the features contain any information on the class-labels. \\
\textbf{Scenario S2: Cyclic data:} We follow \cite{yeung2001model} and generate synthetic data modelling a sinusoidal cyclic behaviour of gene expressions over time. We simulate $n=235$ genes under $D^{(0)}=24$ experiments in $K=10$ classes. Two genes belong to the same class if they have similar phase shifts. 
Each element of the data matrix $y_{ij}$ with $i=1,\dots,n$ and $j=1,\dots,D^{(0)}$ is simulated as
$$y_{ij}=\delta_j+\lambda_j\left[ \alpha_i+\beta_i\sin\left(\frac{2\pi j}{8}-\omega_k+\epsilon_{ij}\right)\right] ,$$ where $\alpha_i\sim\ND(0,1)$ represents the average expression level of gene $i$, $\beta_i\sim\mathcal{N}(3,0.5)$ is the amplitude control of gene $i$, $\lambda_j\sim\mathcal{N}(3,0.5)$ models the amplitude control of condition $j$, while the additive experimental error $\delta_j$ and the idiosyncratic noise $\epsilon_{ij}$ are  drawn independently from $\mathcal{N}(0,1)$ distributions.  The different classes are represented by  $\omega_k$, $k=1,\ldots,K$, which are assumed to be uniformly distributed in $[0,2\pi]$, such that $\omega_k\sim\UD(0,2\pi)$. % from a continuous uniform distribution. 
The sizes of the classes are generated according to Zipf's law, 
meaning that gene $i$ is in class $k$ with probability proportional to $k^{-1}$.
Each observation vector $y_i=(y_{i1},\ldots,y_{i D^{(0)}})^\top$ is individually standardized to have zero mean and unit variance.
Recovering the class labels from the data can be difficult, because the classes are very unbalanced and some classes are very small. On average, only eight observations belong to cluster $k=10$.  Furthermore, it is hard to distinguish between classes $k$ and $k'$ if $\omega_k$ is close to $\omega_{k'}$.\\
\textbf{Wine data}: The dataset from the R-package \texttt{whitening}~\citep{whitening} describes 27 properties of 178 samples of wine from three grape varieties (59 Barolo; 71 Grignolino; 48 Barbera) as reported in \cite{forina1986wine}. \\
\textbf{Olive data}: The dataset from the R-package \texttt{cepp}~\citep{cepp} contains  percentage composition of eight fatty acids found by lipid fraction of 572 Italian olive oils from three regions (323 Southern Italy; 98 Sardinia; 151 Northern Italy), see \cite{forina1983olive}.\\
\textbf{Ecoli data}: This dataset from \citep{DuaGra2019} describes the amino acid sequences of 336 proteins using seven features provided by \cite{ecoliData}. The class is the localization site (143 cytoplasm; 77 inner membrane without signal sequence; 52 perisplasm; 35 inner membrane, uncleavable signal sequence; 20 outer membrane; five outer membrane lipoprotein; two inner membrane lipoprotein; two inner membrane; cleavable signal sequence). \\
\textbf{Vehicle data}: This dataset from the R-package \texttt{mlbench} \citep{leisch2010machine} describes the silhouette of one of four types of vehicles, using a set of 19 features extracted from the silhouette. It consists of 846 observations (218 double decker bus; 199 Cherverolet van; 217 Saab 9000; 212 Opel manta 400). \\
\textbf{Satellite data}: This dataset from \citep{DuaGra2019} consists of four  digital images of the same scene in different spectral bands structured into a $3\times3$ square of pixels defining the neighborhood structure. There are 36 pixels which define the features. We use all 6435 scenes (1533 red soil; 703 cotton crop; 1358 greysoil; 626 damp grey soil; 707 soil with vegetation stubble; 1508 very damp grey soil) as observations. \\
All real datasets are mean-centered and componentwise scaled to have unit variance.

\subsection{Results}
\begin{figure}[ht]
\center
\includegraphics[width=0.94\columnwidth,keepaspectratio]{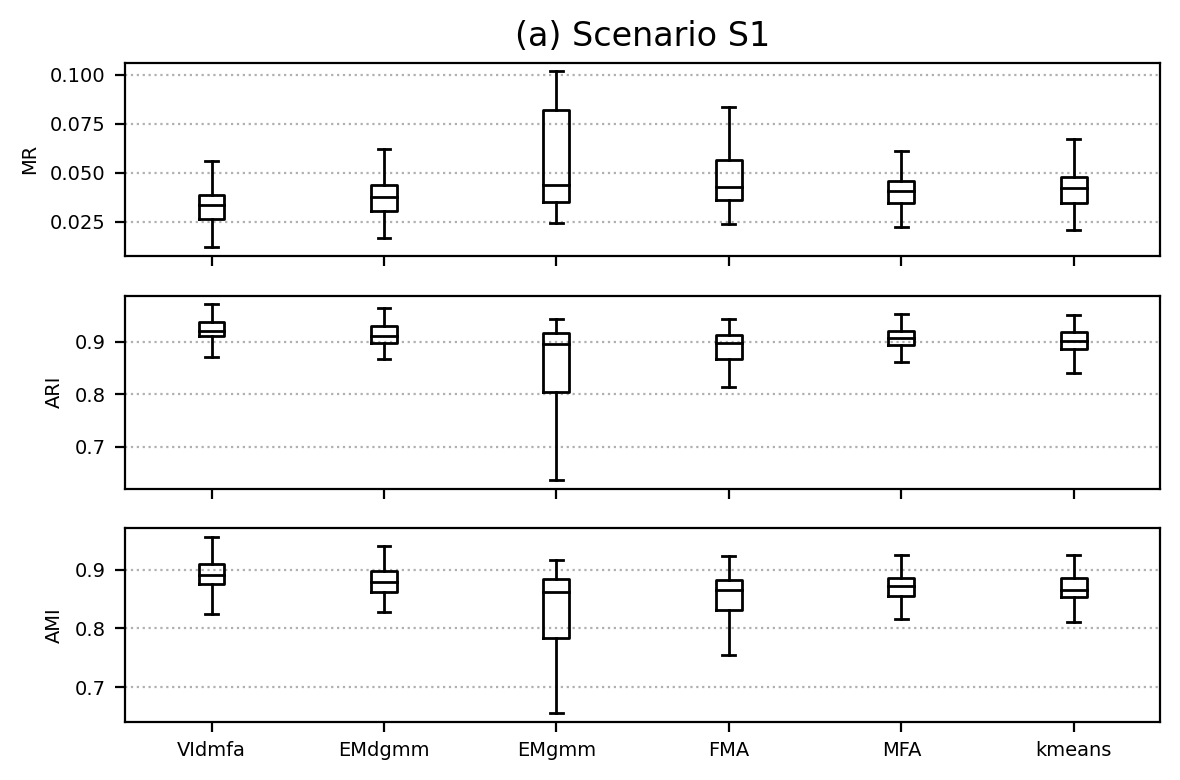}
\includegraphics[width=0.94\columnwidth,keepaspectratio]{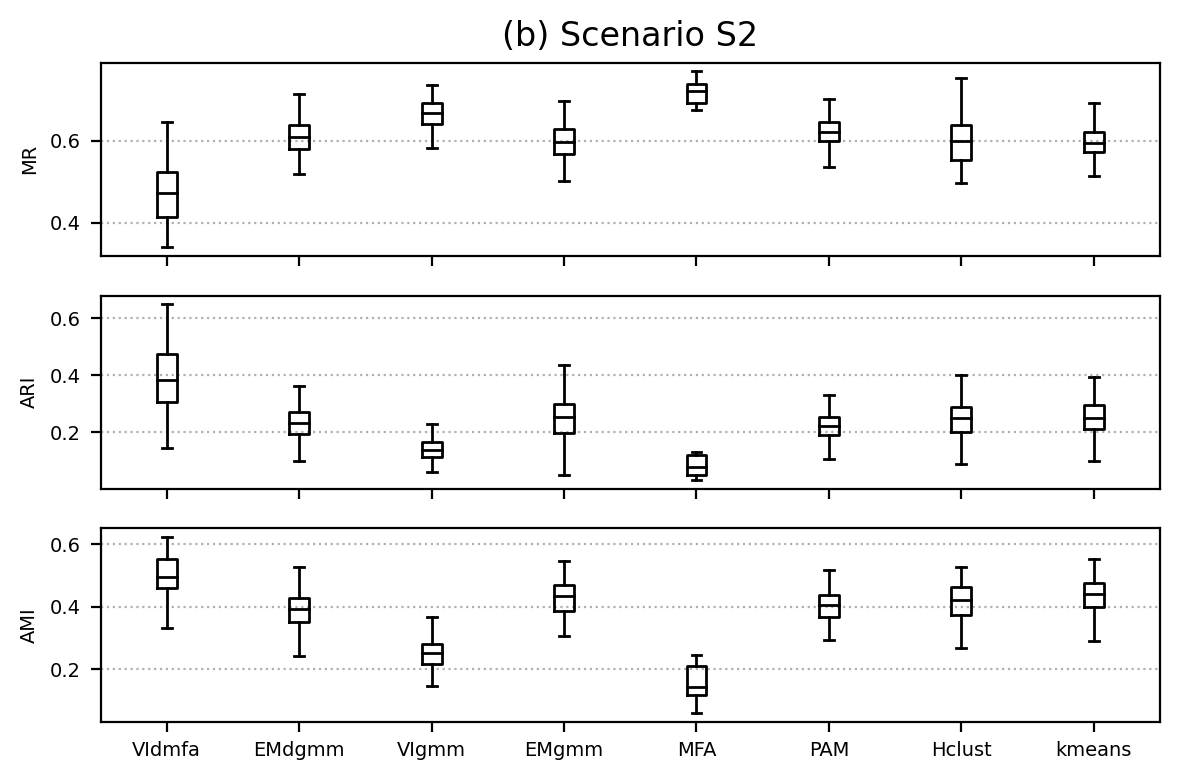}
\caption{\small Scenario S1 (a) and Scenario S2 (b). Boxplots summarize row-wise the MR, ARI across $100$ replications. Here, we present only the best performing methods to make the plots more informative. A plot containing all benchmark methods and runs outside the interquartile range can be found in the Web Appendix {C.3}.}
\label{fig: boxplots}
\end{figure}

\begin{figure}
\center
\includegraphics[width=0.95\columnwidth]{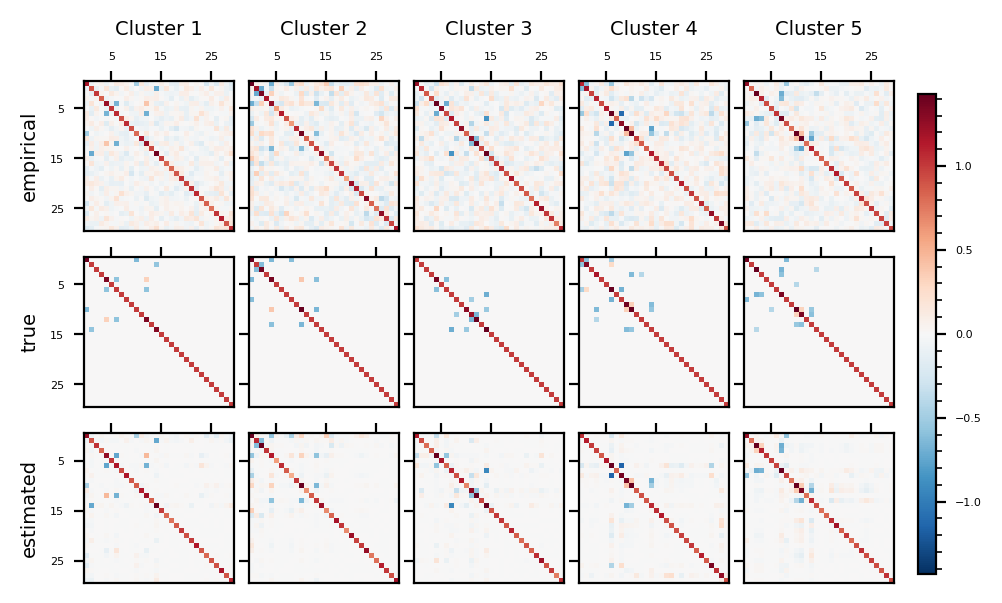}
\caption{\small Scenario S1. Heat-maps of the true, the empirical and the estimated covariance matrices for each of the $5$ clusters. For the calculation of the empirical covariance matrices the true labels of the data points were used, which are not available in practice.}
\label{fig: sparseCovariance}
\end{figure}

To compare the overall clustering performance we assume in a first step that the true number of clusters is known and fixed for all methods. Data-driven choice of the number of clusters is considered later.  For Scenarios S1 and S2, $R=100$ datasets were independently generated. Boxplots of the ARI, AMI and MR across replicates are presented in Figure~\ref{fig: boxplots} for S1 and  S2. Our method outperforms other approaches in both simulated scenarios. Even though the data generating process in Scenario S1 is a GMM, VIdmfa outperforms the classical GMM. One reason for this is that the data simulated in  S1 is relatively noisy, and all information regarding the clusters is contained in a subspace smaller than the number of features. This is not only an assumption for VIdmfa, but also for FMA and MFA. However, compared to the latter two, VIdmfa is able to better recover the sparsity of the covariance matrices due to the shrinkage prior as Figure $\ref{fig: sparseCovariance}$ demonstrates. The clusters derived by VIdmfa are Gaussian in this scenario, because the deeper layer has only one component.
The data generating process of Scenario S2 is non-Gaussian and so are the clusters derived by VIdmfa, due to the fact that the deeper layer might have multiple components.  The performance on the  real data examples is summarized in Table~\ref{tab: benchmarksViroli}. Here our approach is competitive with the other methods considered, but does not outperform the EMdgmm approach.

\begin{sidewaystable}
\begin{adjustbox}{width=0.9\columnwidth,center}
\begin{tabular}{l|ccc|ccc|ccc|ccc|ccc}
       & \multicolumn{3}{c|}{Wine}                        & \multicolumn{3}{c|}{Olive}           & \multicolumn{3}{c|}{Ecoli}                      & \multicolumn{3}{c|}{Vehicle}                     & \multicolumn{3}{c}{Satellite} \\
       & MR             & ARI            & AMI            & MR         & ARI        & AMI        & MR             & ARI            & AMI           & MR             & ARI            & AMI            & MR       & ARI      & AMI      \\ \hline
kmeans & 0.022          & 0.930           & 0.900            & 0.234      & 0.448      & 0.584      & 0.351          & 0.508          & 0.626         & 0.642          & 0.076          & 0.112          & 0.321    & 0.530     & 0.612    \\
PAM    & 0.045          & 0.863          & 0.818          & 0.107      & 0.725      & 0.673      & 0.333          & 0.507          & 0.608         & 0.628          & 0.073          & 0.097          & 0.313    & 0.531    & 0.608    \\
Hclust & 0.045          & 0.865          & 0.845          & 0.386      & 0.493      & 0.671      & 0.333          & 0.518          & 0.624         & 0.634          & 0.092          & 0.114          & 0.397    & 0.446    & 0.577    \\
EMgmm  & 0.011          & 0.964          & 0.953          & 0.348      & 0.524      & 0.68       & 0.259          & 0.676          & 0.635         & 0.599          & 0.095          & 0.167          & 0.414    & 0.465    & 0.557    \\
SNmm   & 0.511          & 0.048          & 0.047          & 0.108      & 0.854      & 0.769      & -              & -              & -             & 0.506          & 0.224          & 0.371          & 0.435    & 0.414    & 0.535    \\
STmm   & 0.427          & 0.194          & 0.218          & 0.098      & 0.864      & 0.785      & 0.298          & 0.594          & 0.571         & \textbf{0.476} & 0.207 & 0.314          & 0.390     & 0.468    & 0.543    \\
FMA    & 0.011          & 0.965          & 0.946          & 0.495      & 0.235      & 0.406      & 0.411          & 0.44           & 0.485         & -              & -              & -              & 0.439        & 0.435        & 0.516        \\
MFA    & 0.006          & 0.983          & 0.973          & 0.052      & 0.914      & 0.855      & -              & -              & -             & 0.573          & 0.153          & 0.201          & \textbf{0.226}    & 0.622    & \textbf{0.666}    \\
VIgmm  & 0.494          & 0.074          & 0.055          & \textbf{0.000} & \textbf{1.000} & \textbf{1.000} & 0.357          & 0.478          & 0.476         & 0.522          & 0.198          & 0.286          & 0.460     & 0.445    & 0.525    \\
EMdgmm & \textbf{0.006} & \textbf{0.982} & 0.936          & \textbf{0.000} & \textbf{1.000} & \textbf{1.000} & \textbf{0.173} & \textbf{0.765} & \textbf{0.710} & 0.506          & 0.204          & 0.238          & 0.246    & \textbf{0.624}    & 0.635    \\
VIdmfa & \textbf{0.006} & \textbf{0.982} & \textbf{0.973} & \textbf{0.000} & \textbf{1.000} & \textbf{1.000} & 0.202          & 0.726          & 0.688         & 0.499          & \textbf{0.208}          & \textbf{0.354} & 0.390        & 0.449        & 0.546       
\end{tabular}
\end{adjustbox}
\caption{\label{tab: benchmarksViroli} \small The MR, ARI and AMI for the real datasets rounded to three decimals is given and the best result is marked in each column. For each method the best out of $10$ runs (according either to BIC or the highest ELBO) is presented.}
\end{sidewaystable}

\begin{sidewaysfigure}
\center
\includegraphics[height=0.7\columnwidth,keepaspectratio]{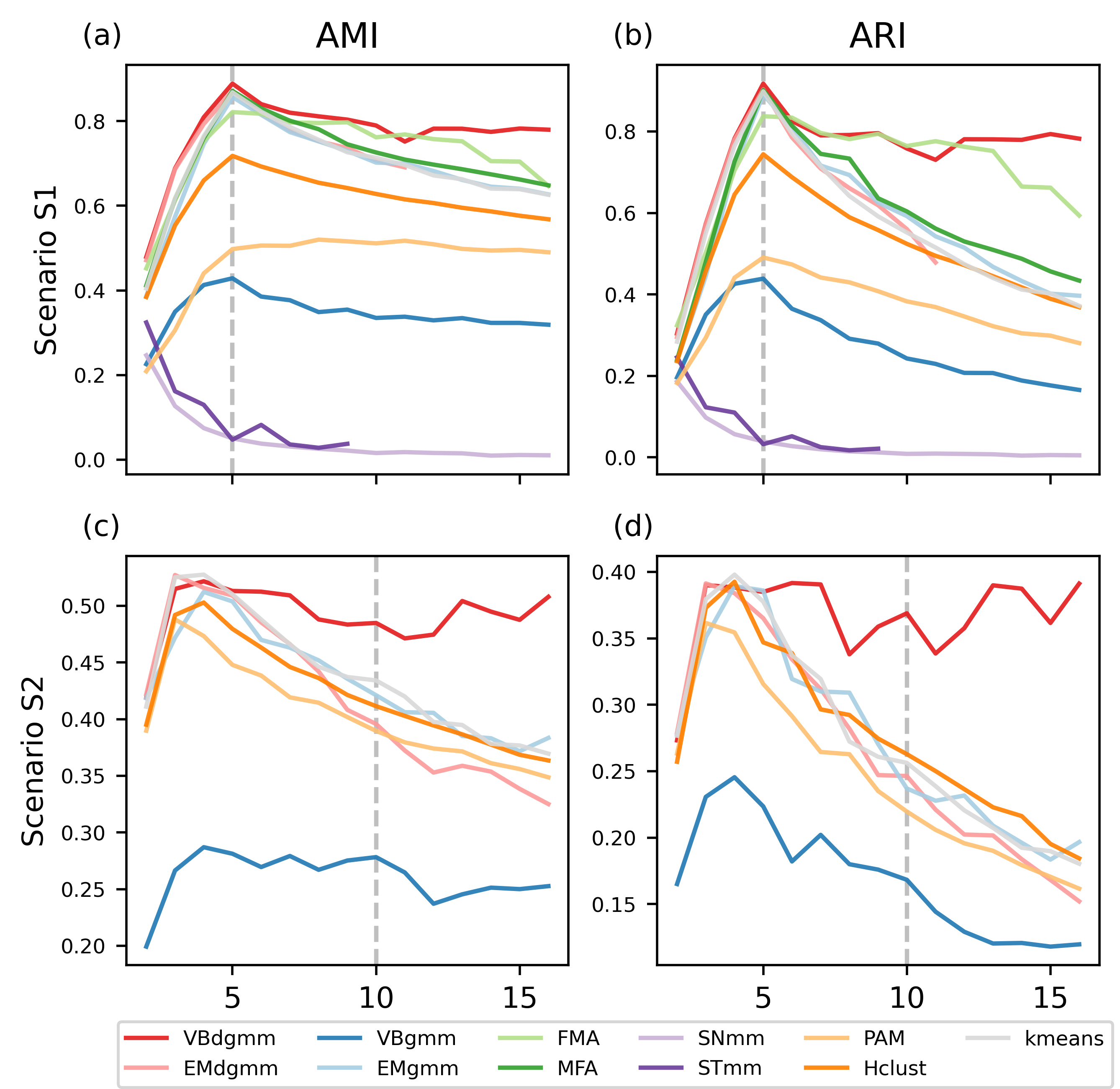}
\caption{\small Scenario S1 (a, b) and Scenario S2 (c, d). The average AMI (a, c) and ARI (b, d) across $10$ independent replicates for different choices on the initial number of clusters $K=2,\dots,16$. The x-axis denotes the initialized number of clusters and the y-axis the average AMI/ARI. The optimal number of clusters for each data set is denoted by the dashed line.}
\label{fig: unknown K}
\end{sidewaysfigure}

The assumption that the number of clusters is known is an artificial one. Clustering is often the starting point of the data investigation process and used to discover structure in the data. Selecting the number of clusters in those settings is sometimes a difficult task. In 
Bayesian mixture modelling this difficulty can be overcome by initializing the model with a relatively
large number of components and using a shrinkage prior on the component weights, so that unnecessary components empty out \citep{rousseau2011asymptotic}. This is also the idea of our model selection process described in Section \ref{sec: model selection}. To indicate that this is also a valid approach for VIdmfa, we fit both Scenarios S1, S2 for different choices of $K^{(0)}$ ranging from $2$ to $16$ and compare the average ARI and AMI over $10$ independent replicates. The results can be found in Figure~\ref{fig: unknown K}. This suggests that our method is able to find the general structure of the data even when $K^{(0)}$ is larger than necessary, but the best results are derived when the model is initialized with the correct number of components. We have found it useful to fit the model with a potentially large number of components and then to refit the model with the number of components selected. This observation justifies our approach to selecting the number of components for the deeper layers discussed in Section~\ref{sec: model selection}. In both real data examples the optimal model architecture is unknown and VIdmfa selects a reasonable model. 

%\FloatBarrier
\section{Application to Gene Expression Data}\label{sec:gene}
In microarray experiments levels of gene expression for a large number of genes are measured under different experimental conditions. In
this example we consider a time course experiment where the data can be represented by  a real-valued expression matrix
$\lbrace m_{ij}$, $1\leq i\leq n$, $1\leq j\leq D\rbrace$, where $m_{ij}$ is the expression level of gene $i$ at time $j$.  We will consider rows of this matrix
to be observations, so that we are clustering the genes according to the time series of their expression values.

Our model is well-suited to the analysis of gene expression datasets.  In many time course microarray experiments both the number
of genes and times is large, and our VIdmfa method scales well with both the sample size and dimension.  
If the time series expression profiles are smooth, their dependence may be well represented using low-rank factor structure.  
Our VIdmfa method also provides a computationally efficient mechanism for
the choice of the number of mixture components. In the simulated Scenario S2 of Section~\ref{sec:Simul}, VIdmfa is able to detect unbalanced and comparable small clusters, which is the main advantage of our VIdmfa approach in comparison to the benchmark methods on this simulated gene expression data set and a similar behaviour is expected in real data applications.

As an example we consider the Yeast data set from  \citet{yeung2001model}, originally considered in \citet{cho1998genome}. This data set has been analyzed by many previous authors \citep{medvedovic2002bayesian,tamayo1999interpreting,lukashin2001analysis} and contains $n=384$ genes over two cell cycles ($D=17$ time points) whose expression levels peak at different phases of the cell cycle. The goal is to cluster the genes according to those peaks and the structure is similar to the simulated cyclic data in Scenario S2 from Section~\ref{sec:Simul}. The genes were assigned to five cell cycle phases by \citet{cho1998genome}, but there are other cluster assignments with more groups available \citep{medvedovic2002bayesian, lukashin2001analysis}. Hence, the optimal number of clusters for this data set is unknown and we aim to illustrate that our VIdmfa is able to select the number of clusters meaningfully. Also, the gene expression levels change relatively quickly between time points in this data set, suggesting a sparse correlation structure. VIdmfa is well suited for this scenario due to the sparsity inducing priors. 

\begin{figure}[ht] 
\centering
\includegraphics[width=0.95\columnwidth,keepaspectratio]{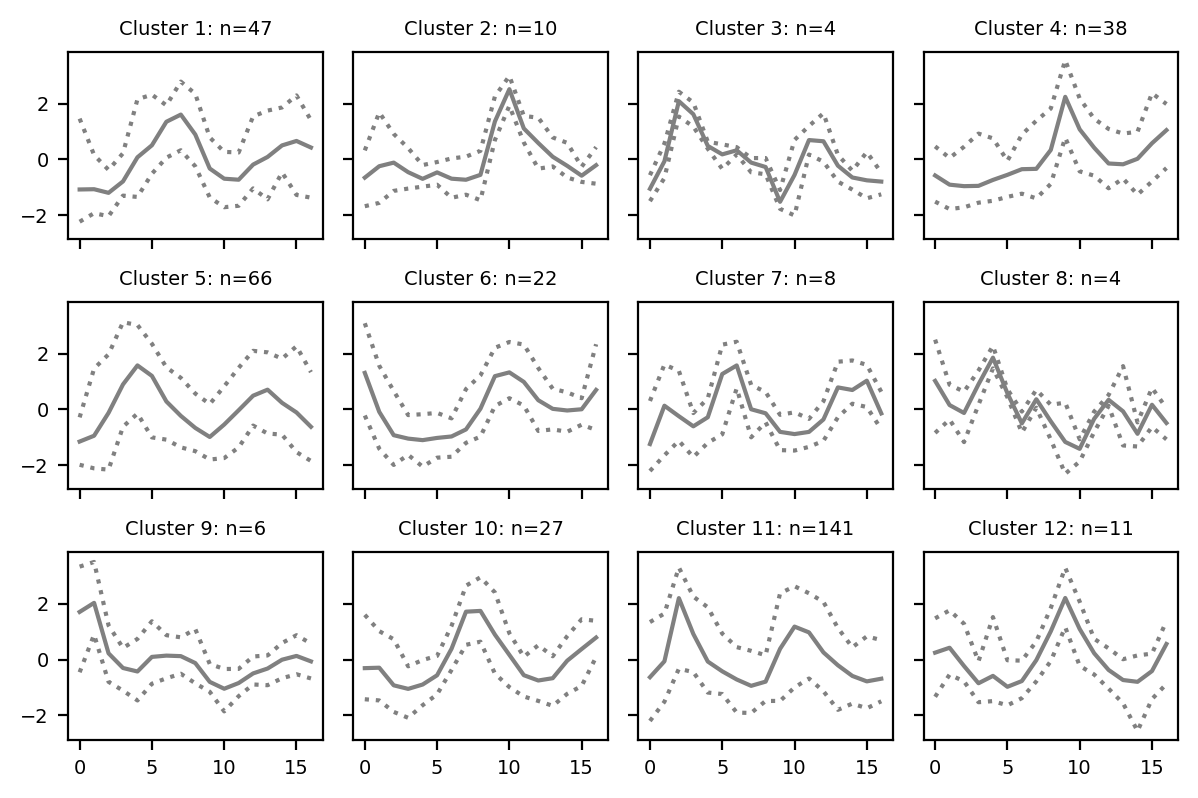} 
\caption{\small Yeast data. VIdmfa selects twelve clusters, when initialized with $K^{(0)}=19$. For each cluster the cluster mean (bold line), as well as minimum and maximum (dotted lines) on the normalized expression patterns are presented and $n$ denotes the number of elements in the respective cluster. The x-axis denotes the time  and the y-axis labels the normalized gene expression.}
\label{fig: clusters yeast}
\end{figure}

As in Scenario S2 and following \citet{yeung2001model}, we individually scale each gene to have zero mean  and unit variance. The simulation study in Section~\ref{sec:Simul} suggests that VIdmfa can be initialized with a large number of potential components when the number of clusters is unknown, since the unnecessary components empty out. 
When initialized with $K^{(0)}=\lfloor\sqrt{n}\rfloor=19$ potential components, the model selection process described in Section~\ref{sec: model selection} selects a model with $L=2$ layers with dimensions $D^{(1)}=4, D^{(2)}=1$ and $K^{(2)}=1$ components in the second layer.
VIdmfa returns twelve clusters as shown in Figure~\ref{fig: clusters yeast}. While some of the clusters are small, others  are large and match well with the clusters proposed in \citet{cho1998genome}. For example, cluster eleven corresponds to their largest cluster.
Comparisons to other kinds of information would be needed 
to decide if all of the twelve clusters are useful here, or if some of the clusters should be merged. For example, clusters two and twelve peak at similar phases of the cell cycle and could possibly be merged, while cluster nine does not seem to have a similar cluster and could be kept even though it contains only $n_9=6$ genes. 
An investigation of the fitted DMFA suggests that the clusters differ mainly by their means, while the covariance matrices have high sparsity, which matches with our expectation of (distant) time points being only weakly correlated. 
\FloatBarrier
\section{Application to Taxi Trajectories}\label{sec:taxi}

In this section, we consider a complex high-dimensional clustering problem for taxi trajectories. We use the publicly available data-set of \citet{moreira2013predicting} consisting of all busy trajectories from 01/07/2013 to 30/06/2014 performed by 442 taxis running in the city of Porto. Each taxi reports its coordinates at 15 second intervals.   The dataset is large in dimension and sample size, with an approximate average of $50$ coordinates per trajectory and $1.7$ million observations in total. Understanding this dataset is a difficult task. \cite{kucukelbir2017automatic} propose a subspace clustering, where they project the data into an eleven-dimensional subspace using a extension on probabilistic principal component analysis in a first step and then use a GMM with 30 components to cluster the lower-dimensional data in a second step. This approach is successful in finding hidden structures in the data like busy roads and frequently taken paths. We will illustrate that VIdgmm is a good alternative because it does not require a separate dimension reduction step and is able to fit a GMM with a much larger number of components. Also VIdgmm scales well with the  sample size, due to the sub-sampling approach described in Section \ref{sec: sgd}.

In our analysis we focus on the paths of the trajectories ignoring the temporal dependencies and interpolate the data accordingly. Each trajectory $f_i$ can be viewed as a mapping $f_i(t): \{t_0,\cdots,t_{T_i}\} \to \mathbb{R}^2$ of $T_i+1$ equally spaced time points onto coordinate pairs $(x_{i,t_h},y_{i,t_h})$ for $h=0,\cdots,T_i$ with $50$ time-points on average. We consider the interpolation $\Tilde{f_i}: [0,1]\to \mathbb{R}^2$, where
\begin{align*}
   \Tilde{f_i}(t)&=\sum_{h=0}^{T_i-1}\mathcal{I}{\left\{\frac{L_{i,h}}{L_{i,T_i}}\leq t\leq\frac{L_{i,h+1}}{L_{i,T_i}}\right\}}(t)\left[\left(1-\frac{t-\frac{L_{i,h}}{L_{i,T_i}}}{\frac{L_{i,h+1}}{L_{i,T_i}}-\frac{L_{i,h}}{L_{i,T_i}}}\right)f_i(t_h)+\frac{t-\frac{L_{i,h}}{L_{i,T_i}}}{\frac{L_{i,h+1}}{L_{i,T_i}}-\frac{L_{i,h}}{L_{i,T_i}}}f_i(t_{h+1})\right], 
\end{align*}
and $L_{i,h}=\sum_{s=0}^{h-1}\vert\vert f_i(t_s)-f_i(t_{s+1})\vert\vert$ is the length of the trajectory up to time-point $t_h$. This interpolation is time-independent in the sense that $d(\Tilde{f_i}(t),\Tilde{f_i}(t+\delta))$ is linear in $t$ for all $\delta>0$, where $d_i(\cdot,\cdot)$ denotes the distance between two points along a linear interpolation of the trajectory $f_i$. This way two trajectories $f_i$ and $f_{i'}$ with the same path, i.e.~the same image in $\mathbb{R}^2$, have similar interpolations $\Tilde{f_i}\approx\Tilde{f_{i'}}$ even when the taxi on one trajectory was much slower than the taxi on the other trajectory.  %which is not true for the original mappings $f_i$ and $f_{i'}$.
Additionally, we consider a tour and its time reversal as equivalent, and therefore change an observation to its time reversal if the origin point after reordering is closer to the city center. We use a discretization of $\Tilde{f_i}(t)$ at $50$ equally spaced points $t=0,\frac{1}{49},\frac{2}{49},\dots,1$ as feature inputs for our clustering algorithm leading to data points in $\mathbb{R}^{100}$. To avoid numerical errors, all coordinates are centered around the city center and scaled to have unit variance.

\begin{sidewaysfigure}[ht]
\centering
\includegraphics[width=0.95\columnwidth,keepaspectratio]{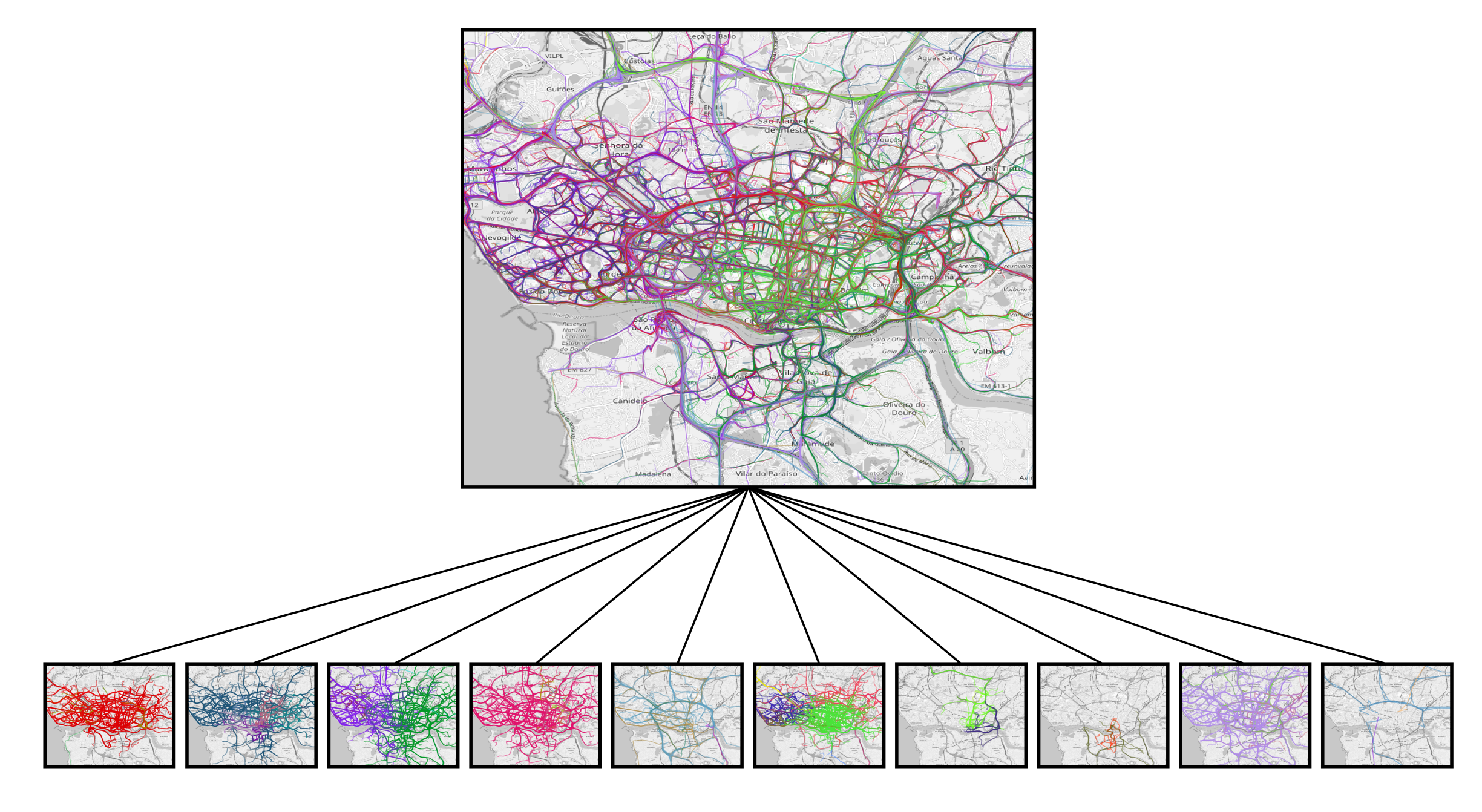}
\caption{\small Taxi data. The nested clustering for $10000$ randomly selected trajectories is shown. There are ten non-empty components in the first layer and nine non-empty components in the second layer, corresponding to a GMM with $90$ components shown in the first row. These clusters can be divided into ten subsets (second row) containing up to nine clusters each. The ten subsets are shown in the second row. Note, that many trajectories connect to endpoints outside the shown region. The map was taken from \texttt{OpenStreetMap}.} %\citep{OpenStreetMap}
\label{fig: taxis_new}
\end{sidewaysfigure}

The architecture chosen has $L=2$ layers with ten components in both layers. The dimensions are set to $D^{(1)}=5$ and $D^{(2)}=2$. This model is equivalent to a $100$-dimensional GMM with $100$ components. %In difference to the application on gene data in Section~\ref{sec:gene} 
We consider each of these Gaussian components as a separate cluster.  In fact, each observation can be matched to one path $k=(k_1,k_2)$ through the model, building Gaussian clusters. 
This idea of linking the different paths to the clusters leads to a natural hierarchical clustering, where the clusters are built layer-wise. First, the observations are divided into $K^{(1)}=10$ large main-clusters based on the components of the first layer, where an observation is in cluster $k_1$ if and only if $\gamma_{ik_1}^{(1)}=1$. Those clusters are then divided further into $K^{(2)}=10$  sub-clusters. An observation of the main cluster $k_1$ is in sub-cluster $k_2$ if $\gamma_{ik_{1}}^{(1)}=1$ and $\gamma_{ik_{2}}^{(2)}=1$, leading to a potential of $K^{(1)}\times K^{(2)}=100$ clusters in total. A graphical representation of this idea based on our fitted DGMM can be found in Figure~\ref{fig: taxis_new}. 

\begin{figure}[htbp] 
\centering
\includegraphics[width=0.95\columnwidth,keepaspectratio]{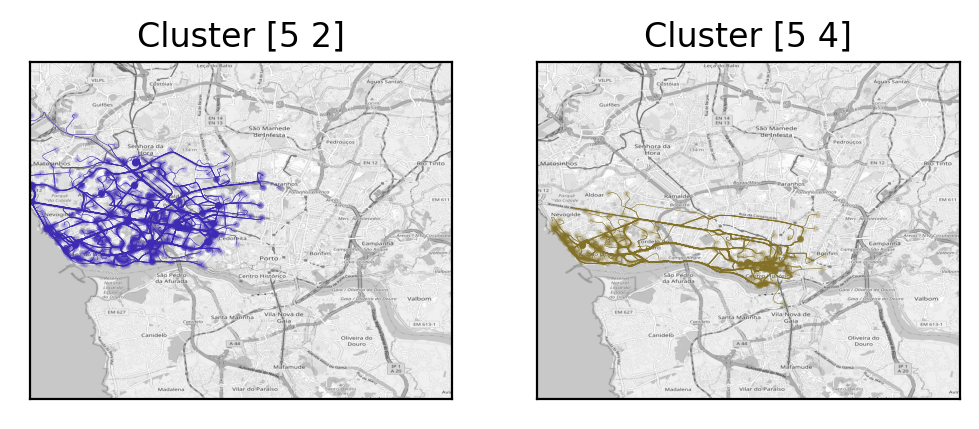} 
\caption{\small Taxi data. VIdmfa detects two clusters consisting of short trajectories connected to the harbour and seaside. Both clusters belong to the same main cluster $5$. Cluster $[5,2]$ connects to a larger region on the west side of the city outside the center with the taxis passing through smaller streets, while cluster $[5,4]$ connects to the city center. Here the Taxis pass through one of three parallels. The dots denote starting and end points of the trajectories.}
\label{fig: taxis_analysis}
\end{figure}

Even though this hierarchical representation has no very clear interpretation for all clusters, it might be used as a starting point for further exploration.
City planners and urban decision makers could be interested in a comparison between start and end points of trajectories in the various clusters to further evaluate the need of connections, for example via public transport, between different regions of the city. Here, the nested structure of the clusters might be helpful, when scanning the clusters for interesting patterns. The $10$ large main clusters based on the first layer give a broad overview, while the sub-clusters allow for a more nuanced analysis. An example is given in Figure~\ref{fig: taxis_analysis}.

%\FloatBarrier
\section{Conclusion and Discussion}\label{sec:conclusion}
In this paper, we introduced a new method for high-dimensional clustering. The method uses sparsity-inducing priors to regularize
the estimation of the DMFA model of \cite{viroli2019deep}, and uses variational inference methods for computation to achieve scalability to large datasets.  We consider the use of overfitted mixtures and ELBO values from short runs to choose a suitable architecture
in a computationally thrifty way.  

As noted recently by \cite{selosse2020bumpy} deep latent variable models like the DMFA are challenging to fit.  It is difficult to estimate
large numbers of latent variables, and there are many local modes in the likelihood function.  
While our sparsity priors are helpful
in some cases for obtaining improved clustering and making the estimation more reliable, 
there is much more work to be done in understanding and robustifying the training of these models.
The way that parameters for mixture components at different layers are combined by considering all paths through the network
allows a large number of mixture components without a large number of parameters, but this feature may also result in probability
mass being put in regions where there is no data.  
It is also not easy to interpret how the deeper layers of these models assist in explaining 
variation.  A deeper understanding of these issues could lead to further improved architectures and interesting new applications in high-dimensional density estimation and regression.

\clearpage
\appendix
\counterwithin{figure}{section} \renewcommand\thefigure{\thesection.\arabic{figure}} 
\renewcommand\thetable{\thesection.\arabic{table}} 
\noindent

\newpage
\singlespacing
\bibliography{litliste}
%\newpage
%\input{tabs}
%\newpage
%\input{figs}
\end{document}

\setlength{\abovedisplayskip}{0.15cm}
\setlength{\belowdisplayskip}{0.15cm}

\pagestyle{empty}
%\singlespacing
%\doublespacing
\begin{center}
\LARGE \textbf{Supporting Information for}\\
\doublespacing
{\LARGE
``Variational Inference and Sparsity in High-Dimensional Deep Gaussian Mixture Models''

}
\end{center}
\tableofcontents
\addtocontents{toc}{\protect\thispagestyle{empty}}
\pagenumbering{gobble}
\newpage
\renewcommand{\thesection}{\Alph{section}}
\setcounter{equation}{0}
\renewcommand{\theequation}{\arabic{equation}}
\section{Variational Approximation and Explicit Derivation of the ELBO}
In this section we will first give the parametric form of the varational density $q_\lambda(\theta)$ and then derive a closed form representation of the evidence lower bound $\calL(\lambda)$.
\subsection{Mean field variational approximation for DMFA}
We consider
the factorized form 
\begin{align} \label{eq:mfvb}
q_\lambda(\theta) & = q(\mu)q(\vec(B))q(z)q(p)q(\delta)q(\gamma)q(\psi)q(g)q(h)q(c)q(\tau)q(\xi).
\end{align}
and each factor in Eq.~\eqref{eq:mfvb} will also be fully factorized.  For the DMFA model, the existence of conditional conjugacy through our prior choices
means that the parametric form
of the factors follows from the factorization assumptions made.  The factors in \eqref{eq:mfvb} are
\begingroup
\allowdisplaybreaks
\begin{align*}
& q(\mu)=\prod_{l=1}^L\prod_{k=1}^{\Kl}\prod_{j=1}^{\Dlm1} q(\mu_{kj}^{(l)}),\;\;\; q(\mu_{kj}^{(l)})=\phi(\mu_{kj}^{(l)};m_{kj}^{(l)}(\mu),\sigma_{kj}^{(l)}(\mu)^2), \\
& q(B)=\prod_{l=1}^L\prod_{k=1}^{\Kl}\prod_{j=1}^{\kapl} q( \vec(B_{k}^{(l)})_j),\;\;\; q(\vec(B_k^{(l)})_j)=\phi(\vec(B_k^{(l)})_j;m_{kj}^{(l)}(B),\sigma_{kj}^{(l)}(B)^2), \\
& q(z)=\prod_{i=1}^n\prod_{l=1}^L \prod_{j=1}^{\Dl} q(z_{ij}^{(l)}),\;\;\; q(z_{ij}^{(l)})=\phi(z_{ij}^{(l)};m_{ij}^{(l)}(z),\sigma_{ij}^{(l)}(z)^2), \\
& q(p)=\prod_{l=1}^L q(p^{(l)})\;\;\;q(p^{(l)})=p_{\Dir}(p^{(l)};d_1^{(l)},\dots, d_{\Kl}^{(l)}), \\
& q(\delta)=\prod_{l=1}^L\prod_{k=1}^{\Kl}\prod_{j=1}^{\Dlm1}q(\delta_{kj}^{(l)}),\;\;\;q(\delta_{kj}^{(l)})=p_{\IGD}(\delta_{kj}^{(l)};a_{kj}^{(l)}(\delta),b_{kj}^{(l)}(\delta)), \\
& q(\gamma)=\prod_{i=1}^n\prod_{l=1}^L q(\gamma_i^{(l)}),\;\;\;q(\gamma_i^{(l)})=p_{\MD}(\gamma_i^{(l)};1,\alpha_{i1}^{(l)},\dots, \alpha_{i\Kl}^{(l)})\\
& q(\psi)=\prod_{l=1}^L \prod_{k=1}^{\Kl}\prod_{j=1}^{\Dlm1} q(\psi_{kj}^{(l)}),\;\;\;q(\psi_{kj}^{(l)})=p_{\IGD}(\psi_{kj}^{(l)};a_{kj}^{(l)}(\psi),b_{kj}^{(l)}(\psi)) \\
&   q(g)=\prod_{l=1}^L\prod_{k=1}^{\Kl}\prod_{j=1}^{\Dlm1}q(g_{kj}^{(l)}),\;\;\;q(g_{kj}^{(l)})=p_{\IGD}(g_{kj}^{(l)};a_{kj}^{(l)}(g),b_{kj}^{(l)}(g)), \\
&   q(h)=\prod_{l=1}^L\prod_{k=1}^{\Kl}\prod_{j=1}^{\kapl}q(h_{kj}^{(l)}),\;\;\;q(h_{kj}^{(l)})=p_{\GaD}(h_{kj}^{(l)};a_{kj}^{(l)}(h),b_{kj}^{(l)}(h)),\\
&   q(c)=\prod_{l=1}^L\prod_{k=1}^{\Kl}\prod_{j=1}^{\kapl}q(c_{kj}^{(l)}),\;\;\;q(c_{kj}^{(l)})=p_{\GaD}(c_{kj}^{(l)};a_{kj}^{(l)}(c),b_{kj}^{(l)}(c)),\\
&   q(\tau)=\prod_{l=1}^L\prod_{k=1}^{\Kl}q(\tau_{k}^{(l)}),\;\;\;q(\tau_{k}^{(l)})=p_{\IGD}(\tau_{k}^{(l)};a_{k}^{(l)}(\tau),b_{k}^{(l)}(\tau)),\\ 
&   q(\xi)=\prod_{l=1}^L\prod_{k=1}^{\Kl}q(\xi_{k}^{(l)}),\;\;\;q(\xi_{k}^{(l)})=p_{\IGD}(\xi_{k}^{(l)};a_{k}^{(l)}(\xi),b_{k}^{(l)}(\xi)),
\end{align*}
\endgroup
where $\phi(\cdot;\mu,\sigma^2)$ denotes the density of a univariate Gaussian distribution, $p_{\IGD}(\cdot;a,b)$/$p_{\GaD}(\cdot;a,b)$ the densities of  inverse gamma/gamma distributions with shape and scale $a,b$, respectively, $p_{\Dir}(\cdot;d_1,\ldots,d_K)$ is the density of a $K$-dimensional Dirichlet distribution and $p_{\MD}(\cdot;1,\alpha_1,\dots,\alpha_k)$ denotes the density of the multinomial distribution with one trial and  probabilities $\alpha_{1},\dots, \alpha_k$. Finally, this implies that $\lambda$ has dimension $$[\sum_{l=1}^L \Kl(8D^{(l-1)}+6\Dl D^{(l-1)}+n+5)+2n\Dl]$$ and 
the complete variational parameter vector is 
\begin{eqnarray*}
\lambda=&(m(\mu),\sigma(\mu),m(B),\sigma(B),m(z),\sigma(z),d,a(\delta),b(\delta),\alpha,a(\psi),b(\psi),\\& a(g),b(g),a(h),b(h),a(c),b(c),a(\tau),b(\tau),a(\xi),b(\xi)).
\end{eqnarray*}

\subsection{Explicit expression of the ELBO}\label{app: calcElbo}

The evidence lower bound is 
\begin{align}
{\cal L}(\lambda) & = E_q(\log (p(\theta)p(y|\theta)))-E_q(\log (q_\lambda(\theta)))\\
& = T_1-T_2 \nonumber
\end{align}
where $T_1=E_q(\log p((\theta)p(y|\theta)))$, $T_2=E_q(\log (q_{\lambda}(\theta)))$ and $E_q(\cdot)$ denotes the expectation with respect to $q_{\lambda}(\theta)$.

We write $T_1$ as 
\begin{align*}
T_1 & = E_q(\log (p(y,z|\theta\setminus \lbrace z\rbrace )))+E_q(\log (p(\mu|g)))+E_q(\log (p(B|h,\tau)))+E_q(\log (p(\delta|\psi)))+E_q(\log (p(\psi))_+\\
& \;\;\;E_q(\log (p(g)))+E_q(\log (p(h|c)))+E_q(\log (p(p)))+E_q(\log (p(\gamma))) +E_q(\log (p(c)))+\\
& \;\;\;E_q(\log (p(\tau|\xi)))+E_q(\log (p(\xi)))\\
& = T_{11}+T_{12}+T_{13}+T_{14}+T_{15}+T_{16}+T_{17}+T_{18}+T_{19}+T_{110}+T_{111}+T_{112}.
\end{align*}
The first term is
\begin{align*}
T_{11} & = E_q(\log p(y,z|\theta\setminus \lbrace z\rbrace)) \\
& = E_q\left(\log \left(\prod_{i=1}^n\prod_{l=1}^L\prod_{k=1}^{\Kl} p(z_i^{(l-1)}|\theta\setminus \{z_i^{(l-1)},\gamma\},\gamma_{ik}^{(l)}=1)^{\gamma_{ik}^{(l)}}\right)\right)+E_q\left(\log \left(\prod_{i=1}^n p(z_i^{(L)})\right)\right) \\
& = \sum_{i=1}^n\sum_{l=1}^L\sum_{k=1}^{\Kl} E_q(\gamma_{ik}^{(l)})E_q(\log (p(z_i^{(l-1)}|\theta\setminus \{z_i^{(l-1)},\gamma\},\gamma_{ik}^{(l)}=1))) +\sum_{i=1}^n E_q(\log (p(z_i^{(L)})))
\end{align*}
where $E_q(\gamma_{ik}^{(l)})=\alpha_{ik}^{(l)}$,
\begin{align*}
E_q(\log p(z_i^{(l-1)}|\theta\setminus \{z_i^{(l-1)},\gamma\},\gamma_{ik}^{(l)}=1) & = 
-\frac{\Dlm1}{2}\log 2\pi-\frac{1}{2}\sum_{j=1}^{\Dlm1} \left\{\log b_{kj}^{(l)}(\delta)-\psi(a_{kj}^{(l)}(\delta))\right\} \\
& \;\;-\frac{1}{2}\frac{a_{kj}^{(l)}(\delta)}{b_{kj}^{(l)}(\delta)}(m_{ij}^{(l-1)}(z)-m_{kj}^{(l)}(\mu)-m_{k[j]}^{(l)}(B)^\top m_{i.}^{(l)}(z))^2 \\
& \;\;+
\sigma_{ij}^{(l-1)}(z)^2+\sigma_{kj}^{(l)}(\mu)^2+\text{Var}_q(\left[B_k^{(l)}z_i^{(l)}\right]_j)
\end{align*}
where $m_{ij}^{(0)}(z)=$ and $\sigma_{ij}^{(0)}(z)^2=0$. %the notation and other expressions appearing here are elaborated further below.  
We have written $m_{k[j]}^{(l)}(B)$ for the mean of the $j$th row of $B_k^{(l)}$, and $m_{i.}^{(l)}(z)$ denotes the vector $(m_{i1}^{(l)}(z),\dots, m_{i\Dl}^{(l)}(z))^\top$.  Also 
\begin{align*}
\text{Var}_q(\left[B_k^{(l)}z_i^{(l)}\right]_j) & = \sum_{r=1}^{\Dl}\left\{m_{k[j]r}^{(l)}(B)^2\sigma_{ir}^{(l)}(z)^2
+m_{ir}^{(l)}(z)^2\sigma_{k[j]r}^{(l)}(B)^2+\sigma_{ir}^{(l)}(z)^2\sigma_{k[j]r}^{(l)}(B)^2\right\}
\end{align*}
where we have written $m_{k[j]r}^{(l)}(B)$ for the mean of element $r$ in the $j$th row of $B_k^{(l)}$, and
$\sigma_{k[j]r}^{(l)}(B)^2$ is the corresponding variance.  Also, 
\begin{align*}
E_q(\log p(z_i^{(L)})) & =  \left\{-\frac{D^{(L)}}{2}\log2\pi-\frac{1}{2}\sum_{j=1}^{D^{(L)}} (m_{ij}^{(L)}(z)^2+\sigma_{ij}^{(L)}(z)^2) \right\}.
\end{align*}
In deriving the above expressions we have used the independence in the variational posterior factorization, and standard results for the expectation of the log and inverse of inverse gamma random variables. 
Continuing with the other terms,
\begin{align*}
T_{12} & = E_q(\log p(\mu|g)) \\
& = \sum_{l=1}^L \sum_{k=1}^{\Kl} \sum_{j=1}^{\Dlm1} \left[-\frac{1}{2}\log 2\pi D^{(l)}
-\frac{1}{2}\left\{\log b_{kj}^{(l)}(g)-\psi(a_{kj}^{(l)}(g))\right\}-\frac{1}{2 G^{(l)}} \frac{a_{kj}^{(l)}(g)}{b_{kj}^{(l)}(g)} 
\left\{\sigma_{kj}^{(l)}(\mu)^2+m_{kj}^{(l)}(\mu)^2\right\}\right],
\end{align*}
\begin{align*}
T_{13} & = E_q(\log p(B|h,\tau)) \\
& = \sum_{l=1}^L \sum_{k=1}^{\Kl} \sum_{j=1}^{\kapl} \bigg[-\frac{1}{2}\log 2\pi
+\frac{1}{2}\left\{\psi(a_{kj}^{(l)}(h))-\log b_{kj}^{(l)}(h)\right\}-\frac{1}{2}\left\{\log b_{k}^{(l)}(\tau)-\psi(a_{k}^{(l)}(\tau))\right\}-\\
&\;\;\;\frac{1}{2} \frac{a_{k}^{(l)}(\tau)a_{kj}^{(l)}(h)}{b_{k}^{(l)}(\tau)b_{kj}^{(l)}(h)} 
\left\{\sigma_{kj}^{(l)}(B)^2+m_{kj}^{(l)}(B)^2\right\}\bigg],
\end{align*}
\begin{align*}
T_{14} & = E_q(\log p(\delta|\psi)) \\
& = \sum_{l=1}^L \sum_{k=1}^{\Kl} \sum_{j=1}^{\Dlm1} \left[
-\frac{1}{2}\left\{\log b_{kj}^{(l)}(\psi)-\psi(a_{kj}^{(l)}(\psi))\right\}
-\log \Gamma\left(\frac{1}{2}\right)
-\frac{3}{2}\left\{\log b_{kj}^{(l)}(\delta)-\psi(a_{kj}^{(l)}(\delta))\right\} \right. \\
%   & \;\;\;\left. -\frac{b_{kj}^{(l)}(\delta)b_{kj}^{(l)}(\psi)}{(a_{kj}^{(l)}(\delta)-1)(a_{kj}^{(l)}(\psi)-1)}\right].
& \;\;\;\left. -\frac{a_{kl}^{(l)}(\delta)a_{kj}^{(l)}(\psi)}{b_{kj}^{(l)}(\delta)b_{kj}^{(l)}(\psi)}\right],
\end{align*}
\begin{align*}
T_{15} & = E_q(\log p(\psi)) \\
& = \sum_{l=1}^L\sum_{k=1}^{\Kl}\sum_{j=1}^{\Dlm1} \left\lbrack -\frac{1}{2} \log({A^{(l)}}^2) -\log \Gamma\left(\frac{1}{2}\right)
-\frac{3}{2}\left\{\log b_{kj}^{(l)}(\psi)-\psi(a_{kj}^{(l)}(\psi)\right\}-\frac{a_{kj}^{(l)}(\psi)}{b_{kj}^{(l)}(\psi)}\frac{1}{A^{(l)}}^2\right\rbrack,
\end{align*}
\begin{align*}
T_{16} & = E_q(\log p(g)) \\
& = \sum_{l=1}^L \sum_{k=1}^{\Kl}\sum_{j=1}^{\Dlm1} \left[\frac{1}{2}\log \frac{1}{2}-\log \Gamma\left(\frac{1}{2}\right)
-\frac{3}{2}\left\{\log b_{kj}^{(l)}(g)-\psi(a_{kj}^{(l)}(g))\right\}-\frac{a_{kj}^{(l)}(g)}{2b_{kj}^{(l)}(g)}\right],
\end{align*}
\begin{align*}
T_{17} & = E_q(\log p(h|c)) \\
& = \sum_{l=1}^L \sum_{k=1}^{\Kl} \sum_{j=1}^{\kapl} \bigg[-\log\left(\Gamma\left(\frac{1}{2}\right)\right)+\frac{1}{2}(\psi(a_{kj}^{(l)}(c))-\log(b_{kj}^{(l)}(c)))-\\
&\;\;\; \frac{1}{2}(\psi(a_{kj}^{(l)}(h))-\log(b_{kj}^{(l)}(h)))-\frac{a_{kj}^{(l)}(c)a_{kj}^{(l)}(h)}{b_{kj}^{(l)}(c)
	b_{kj}^{(l)}(h)}  \bigg],
\end{align*}
\begin{align*}
T_{18} & = E_q(\log p(p)) \\
& = \sum_{l=1}^L\left(-\log \Gamma(\rho_{.}^{(l)}) +\sum_{k=1}^{\Kl} \left[ \log \Gamma (\rho_k^{(l)}) 
+ (\rho_k^{(l)}-1)\left\{\psi(d_k^{(l)})-\psi(d_.^{(l)})\right\}\right]\right),
\end{align*}
where $\rho_.^{(l)}=\sum_k \rho_k^{(l)}$, $d_.^{(l)}=\sum_k d_k^{(l)}$, 
\begin{align*}
T_{19} & = E_q(\log p(\gamma)) \\
& = \sum_{i=1}^n\sum_{l=1}^L \sum_{k=1}^{\Kl} \alpha_{ik}^{(l)}\left\{\psi(d_k^{(l)})-\psi(d_.^{(l)})\right\},
\end{align*}
\begin{align*}
T_{110} & = E_q(\log p(c)) \\
& = \sum_{l=1}^L \sum_{k=1}^{\Kl} \sum_{j=1}^{\kapl} \bigg[-\log\left(\Gamma\left(\frac{1}{2}\right)\right)-\frac{1}{2}(\psi(a_{kj}^{(l)}(c))-\log(b_{kj}^{(l)}(c)))-\frac{a_{kj}^{(l)}(c)}{b_{kj}^{(l)}(c)}  \bigg],
\end{align*}
\begin{align*}
T_{111} & = E_q(\log p(\tau|\xi)) \\
& = \sum_{l=1}^L \sum_{k=1}^{\Kl}  \bigg[-\log\left(\Gamma\left(\frac{1}{2}\right)\right)-\frac{1}{2}(\log(b_{k}^{(l)}(\xi))-\psi(a_{k}^{(l)}(\xi)) )   -\frac{3}{2}(\log(b_{k}^{(l)}(\tau))-\psi(a_{k}^{(l)}(\tau)))-\\
&\;\;\;\frac{a_{k}^{(l)}(\tau)a_{k}^{(l)}(\xi)}{b_{k}^{(l)}(\tau)b_{k}^{(l)}(\xi)}  \bigg],
\end{align*}
and
\begin{align*}
T_{112} & = E_q(\log p(\xi)) \\
& = \sum_{l=1}^L \sum_{k=1}^{\Kl}  \bigg[-\log\left(\Gamma\left(\frac{1}{2}\right)\right)-\frac{1}{2}\log((\nu^{(l)})^2)-\frac{3}{2}(\log(b_{k}^{(l)}(\xi))-\psi(a_{k}^{(l)}(\xi)))-\frac{1}{(\nu^{(l)})^2}\frac{a_{k}^{(l)}(\xi)}{b_{k}^{(l)}(\xi)}  \bigg].
%\frac{b_{k}^{(l)}(\xi)}{a_{k}^{(l)}(\xi)-1}  \bigg].
\end{align*}
We write the second term in the evidence lower bound $T_2=E_q(\log q(\theta))$ as 
\begin{align*}
T_2 & = E_q(\log q_{\lambda}(\mu))+E_q(\log q_{\lambda}(B))+E_q(\log q_{\lambda}(z)) + E_q(\log q_{\lambda}(p))+E_q(\log q_{\lambda}(\delta)) \\
& \;\;\;+E_q(\log q_{\lambda}(\psi))+E_q(\log q_{\lambda}(g))+E_q(\log q_{\lambda}(h))+E_q(\log q_{\lambda}(\gamma))+\\
&\;\;\;E_q(\log q_{\lambda}(c))+E_q(\log q_{\lambda}(\tau))+E_q(\log q_{\lambda}(\xi)) \\
& = T_{21}+T_{22}+T_{23}+T_{24}+T_{25}+T_{26}+T_{27}+T_{28}+T_{29}+T_{210}+T_{211}+T_{212}.
\end{align*}
Expressions for the terms above follow directly from well-known expressions for the entropy for normal, gamma, inverse gamma, Dirichlet and multinomial distributions.  Specifically,
\begin{align*}
T_{21} & = -\frac{1}{2}\sum_{l=1}^L \sum_{k=1}^{\Kl}\sum_{j=1}^{\Dlm1} \log (2\pi e \sigma_{kj}^{(l)}(\mu)^2),
\end{align*}
\begin{align*}
T_{22} & = -\frac{1}{2}\sum_{l=1}^L \sum_{k=1}^{\Kl}\sum_{j=1}^{\kapl} \log (2\pi e \sigma_{kj}^{(l)}(B)^2),
\end{align*}
\begin{align*}
T_{23} & = -\frac{1}{2}\sum_{i=1}^n \sum_{l=1}^L \sum_{j=1}^{\Dl} \log (2\pi e \sigma_{ij}^{(l)}(z)^2),
\end{align*}
\begin{align*}
T_{24} & = \sum_{l=1}^L\left[-\log \mathcal{B}(d)-(d_{.}^{(l)}-\Kl)\psi(d_.^{(l)})+\sum_{k=1}^{\Kl} \left\{(d_k^{(l)}-1)\psi(d_k^{(l)})\right\}\right],
\end{align*}
where $d_.^{(l)}=\sum_k d_k^{(l)}$ and $\mathcal{B}(d)=\prod_{k=1}^{\Kl}\Gamma(d_k^{(l)})/\Gamma(d_.^{(l)})$, 
\begin{align*}
T_{25} & = \sum_{l=1}^L\sum_{k=1}^{\Kl}\sum_{j=1}^{\Dlm1}\left[(1+a_{kj}^{(l)}(\delta))\psi(a_{kj}^{(l)}(\delta))
-a_{kj}^{(l)}(\delta)-\log \left\{b_{kj}^{(l)}(\delta)\Gamma(a_{kj}^{(l)}(\delta))\right\}\right],
\end{align*}
\begin{align*}
T_{26} & = \sum_{l=1}^L\sum_{k=1}^{\Kl}\sum_{j=1}^{\Dlm1} \left[(1+a_{kj}^{(l)}(\psi))\psi(a_{kj}^{(l)}(\psi))
-a_{kj}^{(l)}(\psi)-\log\left\{b_{kj}^{(l)}(\psi)\Gamma(a_{kj}^{(l)}(\psi))\right\}\right],
\end{align*}

\begin{align*}
T_{27} & = \sum_{l=1}^L \sum_{k=1}^{\Kl}\sum_{j=1}^{\Dlm1} \left[(1+a_{kj}^{(l)}(g))\psi(a_{kj}^{(l)}(g))-
a_{kj}^{(l)}(g)-\log\left\{b_{kj}^{(l)}(g)\Gamma(a_{kj}^{(l)}(g))\right\}\right],
\end{align*}
\begin{align*}
T_{28} & = \sum_{l=1}^L\sum_{k=1}^{\Kl}\sum_{j=1}^{\kapl}\left[a_{kj}^{(l)}(h)-\log(b_{kj}^{(l)}(h))+\log(\Gamma(a_{kj}^{(l)}(h)))+(1-a_{kj}^{(l)}(h))\psi(a_{kj}^{(l)}(h))\right],
\end{align*}
\begin{align*}
T_{29} & = \sum_{i=1}^n\sum_{l=1}^L \sum_{k=1}^{\Kl} \alpha_{ik}^{(l)}\log \alpha_{ik}^{(l)},
\end{align*}
\begin{align*}
T_{210} & = \sum_{l=1}^L\sum_{k=1}^{\Kl}\sum_{j=1}^{\kapl}\left[a_{kj}^{(l)}(c)-\log(b_{kj}^{(l)}(c))+\log(\Gamma(a_{kj}^{(l)}(c)))+(1-a_{kj}^{(l)}(c))\psi(a_{kj}^{(l)}(c))\right],
\end{align*}
\begin{align*}
T_{211} & = \sum_{l=1}^L\sum_{k=1}^{\Kl}\sum_{j=1}^{\kapl}\left[(1+a_{kj}^{(l)}(\tau))\psi(a_{kj}^{(l)}(\tau))
-a_{kj}^{(l)}(\tau)-\log \left\{b_{kj}^{(l)}(\tau)\Gamma(a_{kj}^{(l)}(\tau))\right\}\right],
\end{align*}
and
\begin{align*}
T_{212} & = \sum_{l=1}^L\sum_{k=1}^{\Kl}\sum_{j=1}^{\kapl}\left[(1+a_{kj}^{(l)}(\xi))\psi(a_{kj}^{(l)}(\xi))
-a_{kj}^{(l)}(\xi)-\log \left\{b_{kj}^{(l)}(\xi)\Gamma(a_{kj}^{(l)}(\xi))\right\}\right],
\end{align*}

\section{Detailed Description of the Algorithm}

The complete algorithm is given by Algorithm~\ref{algo: complete}. In Section~\ref{app: initalizations} we will derive details on the initialization of $\lambda_G$ and in Section~\ref{app: local optimization} the local optimization step is described in more detail. 

{\setstretch{1.0}
\begin{algorithm}[ht]
\SetAlgoLined
 Initialize $\lambda_G$, $t=0$ as described in Appendix \ref{app: initalizations} \;
 \While{$\overline{\cal L}(\lambda_G)$ is not converged}{
 Draw a mini batch $A$\;
Fix the global model-parameters $\beta$ based on the mode of $q_{\lambda_G}(\beta)$\;
\For{$i\in A$}{
Find the optimal path $\gamma_i=\arg\max_{\gamma_i} p(\gamma_i\vert \beta, y_i)$.\;
set $\alpha_ {ik}^{(l)}=1$ if $\gamma_{ik}^{(l)}=1$ and $\alpha_ {ik}^{(l)}=0$ otherwise\;
\For{$l=1,\dots,L$}{
set $q(z_{ij}^{(l)})\overset{d}{=}p(z_{ij}^{(l)}\vert z_{i}^{(l-1)},\gamma_i,\beta,y_i)$, where $z_{i}^{(0)}=y_{i}$ and $z_{ij}^{(l-1)}=\dsE_q[q(z_{ij}^{(l-1)})]$ for $l>0$\;
}
set $M^{(i)}(\lambda_G)=(m_i(z),\sigma_i(z)^2,\alpha_i)$ the local parameters for observation $i$ \;}
Calculate $I_F(\lambda_G)^{-1}\widehat{\nabla_{\lambda_G}\overline{\calL}}(\lambda_G)$ based on $M^{(i)}(\lambda_G)$ for $i\in A$\;
Update the scalar step size $\alpha_t>0$ as in \cite{ranganath2013adaptive}\;
Update $\lambda_G\leftarrow \lambda_G+a_t I_F(\lambda_G)^{-1} \widehat{\nabla_{\lambda_G}\overline{\cal L}}(\lambda_G),$\;
$t\leftarrow t+1$\;
 }
\Return $\lambda_G$\;
 \caption{Fitting a Bayesian DMFA}
 \label{algo: complete}
\end{algorithm}
}

\subsection {Initialization of the variational optimization}\label{app: initalizations}
The ELBO suffers from local optima. Hence, the starting values for $\lambda$ have to be chosen carefully, since poor starting values will lead to poor local optima. We propose the following procedure.

In a first step we fit the model in a naive way giving estimates for $\mukl,\Bkl,\zil$ and $\epsikl$ for $k=1,\dots,K_l, l=1,\dots,L$ and then fit starting values for $\lambda_G$ based on this naive model fit. 
This is done via a layer-wise approach where for each layer $l=1,\dots,L$ we cluster the data using $K_l$-means into $K_l$ subsets. The cluster mean is a natural estimate for $\mukl$. A principal component analysis with $L1$-penalty gives estimators for $\Bkl$ and $\zil$. We estimate $\epsikl=z_{i}^{(l-1)}-\mukl-\Bkl\zil$ and use the estimates of $\zil$ to fit layer $l+1$. This leads to starting values for $m_{kj}^{(l)}(\mu),\sigma_{kj}^{(l)}(\mu)^2,m_{kj}^{(l)}(B),\sigma_{kj}^{(l)}(B)^2,a_{kj}^{(l)}(\delta),b_{kj}^{(l)}(\delta)$ and $p_k^{(l)}$.
Since the priors are conditionally conjugate, we can orient the remaining starting values for $\lambda_G$ on the full conditional posteriors, which are given by
\begin{align*}
& g_{kj}^{(l)}|\mu_{kj}^{(l)}\sim \IGD\left(\frac{1}{2},\frac{1}{2}+\frac{\mu_{kj}^{(l)}}{2G^{(l)}}\right),\\
& h_{kj}^{(l)}|\vec (\Bkl)_j,c_{kj}^{(l)},\tau_{k}^{(l)}\sim \GaD\left(1,\frac{\vec (\Bkl)_j^2}{2\tau_{k}^{(l)}}+c_{kj}^{(l)}\right),\\
&c_{kj}^{(l)}|\vec (\Bkl)_j,h_{kj}^{(l)}\sim  \GaD\left(\frac{1}{2},1+h_{kj}^{(l)}\right),\\
&\tau_k^{(l)}|\vec (\Bkl),h_{k\cdot}^{(l)},\xi_k^{(l)}\sim \IGD\left(\frac{D^{(l)}\cdot D^{(l+1)}+1}{2},\frac{1}{\xi_k^{(l)}}+\frac{\sum_j h_{kj}^{(l)}\vec (\Bkl)_j}{2}\right)\\
&\xi_k^{(l)}|\tau_k^{(l)}\sim \IGD\left(\frac{1}{2},\frac{1}{\tau_k^{(l)}}+\frac{1}{(\nu^{(l)})^2}\right)\\
&\psi_{kj}^{(l)}|\delta_{kj}^{(l)}\sim \IGD\left(\frac{1}{2},\frac{1}{{A^{(l)}}^2}+\frac{1}{\delta_{kj}^{(l)}}\right).
\end{align*}

This leads to the starting values:
\begin{align*}
&m_{kj}^{(l)}(\mu)=\mu_{kj}^{(l)},\;\;\;\sigma_{kj}^{(l)}(\mu)^2= \frac{\sum_{i=1}(z_i^{(l-1)}-\mukl)^2}{n^2},\\
&m_{kj}^{(l)}(B)=B_{kj}^{(l)},\;\;\;\sigma_{kj}^{(l)}(B)^2=0.001, \\
&d_{\Kl}^{(l)}=p_k^{(l)}n \\
& a_{kj}^{(l)}(\delta)=\frac{p_k^{(l)}n+1}{2},\;\;\;b_{kj}^{(l)}(\delta)=\frac{\sum_{i=1}^n(\epsikl)^2}{2}, \\
& a_{kj}^{(l)}(\psi)=1,\;\;\;b_{kj}^{(l)}(\psi)=\frac{n}{\sum_{i=1}^n(\epsikl)^2}+\frac{1}{(A^{(l)})^2}, \\
& a_{kj}^{(l)}(g)=1,\;\;\;b_{kj}^{(l)}(g)=\frac{(\mukl)^2}{2(G^{(l)}+1)}, \\
& a_{kj}^{(l)}(h)=1,\;\;\;b_{kj}^{(l)}(h)=\frac{\vec (\Bkl)_j^2}{2(\nu^{(l)})^2}+0.5,\\
& a_{kj}^{(l)}(c)=1,\;\;\;b_{kj}^{(l)}(c)=1,\\
& a_{k}^{(l)}(\tau)=\frac{D^{(l)}D^{(l+1)}}{2}\;\;\;,b_{k}^{(l)}(\tau)=\frac{3}{2(\nu^{(l)})^2},\\ 
& a_{k}^{(l)}(\xi)=1,\;\;\;b_{k}^{(l)}(\xi)=D^{(l)}D^{(l+1)}+2+\frac{1}{(\nu^{(l)})^2}.
\end{align*}
If the sample size $n$ is large the initialization can be done using a minibatch.

\subsection{Local optimization step}\label{app: local optimization}

Assume estimates for all global parameters $\beta$ are fixed. Along a path $(k_1,\dots,k_L)$, $z^{(l)}$ is normal for all $l=0,\dots,L$ and the location and scale parameters are given by the following recursion:
\begin{align*}
	z^{(l-1)}\sim \mathcal{N}(m^{(l-1)},V^{(l-1)}),
\end{align*}
where $m^{(l-1)}=\mu_{k_l}^{(l)}+B_{k_l}^{(l)}m_{k_l}^{(l)}$ and $V^{(l-1)}=B_{k_l}^{(l)}V^{(l)}\left(B_{k_l}^{(l)}\right)^\top+\delta_{k_l}^{(l)}$ and $m^{(L)}=0, V^{(L)}=I_{d_L}$.
Hence, the conditional $p(z_{i}^{(l)}\vert z_i^{(l-1)},\gamma_i,\beta)$ is Gaussian and given as
$$p(z_{i}^{(l)}\vert z_i^{(l-1)},\gamma_i,\beta)=\phi\left(z_{i}^{(l)};C\left[\left(B_{k_l}^{(l)}\right)^\top\delta_{k_l}^{(l)^\top}(\mu_{k_l}^{(l)}-z^{(l-1)})-V^{(l)^{-1}}m^{(l)}\right],C\right),$$
where $C=\left(\left(B_{k_l}^{(l)}\right)^\top\delta_{k_l}^{(l)^{-1}}B_{k_l}^{(l)}+V^{(l)^{-1}}\right)^{-1}$, leading directly to a closed form representation of $p(z_{ij}^{(l)}\vert z_i^{(l-1)},\gamma_i,\beta)$. This leads to the layerwise approximation of $M^{(i)}(\lambda_G)$ described in Algorithm~\ref{algo: local step}.
Finding the optimal path $\gamma_i=\arg\max_{\gamma_i} p(\gamma_i\vert \theta_G)$ can be computational difficult if $D^{(0)}$ and the number of components in the layers are very large, but in this case also the dimension of $\lambda_G$ is high, which would lead to a potentially long run time of a gradient ascent algorithm as well.
{\setstretch{1.0}
\begin{algorithm}[ht]
\SetAlgoLined
\For{$i\in A$}{
Find the optimal path $\gamma_i=\arg\max_{\gamma_i} p(\gamma_i\vert \beta, y_i)$\;
set $\alpha_ {ik}^{(l)}=1$ if $\gamma_{ik}^{(l)}=1$ and $\alpha_ {ik}^{(l)}=0$ otherwise\;
\For{$l=1,\dots,L$}{
set $z_{ij}^{(l-1)}=\dsE_q[q(z_{ij}^{(l-1)})]=m^{(l-1)}_{ij}(z)$\;
Calculate $m^{(l-1)},V^{(l-1)}$ and $C$ as described above\;
$m^{(l)}_{ij}(z)=(C(B^{(l)\top}\delta^{(l)^\top}(\mu^{(l)}-z_i^{(l-1)})-V^{(l)^{-1}}m^{(l)}))_j$\;
$\sigma^{(l)}_{ij}(z)=C_{jj}$
}
set $M^{(i)}(\lambda_G)=(m_i(z),\sigma_i(z)^2,\alpha_i)$ the local parameters for observation $i$ \;}
\Return $M^{(i)}(\lambda_G)$ for $i\in A$\;
 \caption{Detailed local step of Algorithm~\ref{algo: complete}}
 \label{algo: local step}
\end{algorithm}
}

\FloatBarrier
\section{Details on the Simulation Study}\label{supp: simul} 

In this section we will give more details on the simulation study. Details on the benchmark methods are given in Section~\ref{supp: bms} and the measures of performance are described in Section~\ref{supp: meassures}. Section~\ref{supp: results} is dedicated to the results, which are not presented in the original paper.

\subsection{Benchmark methods}\label{supp: bms}
We compare our method, which we label \underline{VIdmfa}, to the following benchmark.\\
\underline{EMdgmm}: The EM-based DMFA of \cite{viroli2019deep} is  estimated with the R package  \texttt{deepgmm} \citep{deepgmm}. The model architecture is selected according to the BIC based on all possible models with two or three layers and the number of components for the deeper layers ranging from one to five.  \\
\underline{EMgmm}: A  GMM based on classical EM algorithm using the Python package \texttt{scikit-learn} \citep{scikit-learn}.\\
\underline{VIgmm}: A GMM using the VI approach by \citet{mcgrory2007variational}. \\
\underline{SNmm}: A skew-normal mixture  from the R-package \texttt{EmmixSkew} \citep{EMMIXskew}. \\
\underline{STmm}: A skew-t mixture  from the R-package \texttt{EmmixSkew}  \citep{EMMIXskew}. \\
\underline{kmeans}: A simple k-means algorithm (denoted as \underline{k-means}) from the Python package \texttt{scikit-learn} \citep{scikit-learn}. \\
\underline{PAM}: A partition around metroids  from the R-package \texttt{cluster} \citep{cluster}.\\
\underline{Hclust}: Hierarchical clustering with Ward distance using the Python-package \texttt{scikit-learn} \citep{scikit-learn}.\\
\underline{FMA}: A factor mixture analyser   from the R-package \texttt{FactMixtAnalysis} \citep{FactMixtAnalysis}.\\
\underline{MFA}: A mixture of factor analyzer \citep{mclachlan2003modelling} from the R-package \texttt{EMMIXmfa} \citep{EMMIXmfa}.

\subsection{Measures of performance} \label{supp: meassures}
Each dataset comes with `ground truth' labels we wish to recover. To measure how close a set of clusters is to the ground truth labels, 
we consider three popular performance  measures (e.g.~\cite{vinh2010information}).  These are 
the missclassification rate (MR), the adjusted rand index (ARI) and the adjusted mutual information (AMI). 

The MR is given as the percentage of data points belonging to different clusters under the best possible matching of clusters
and classes. Hence, it takes values between $0$ and $1$, where a $0$ value means that both clusterings are equal (up to label names).
The ARI \citep{hubert1985comparing} counts the pairs of data points on which the ground truth and employed method agree or disagree.  Denote by $N_{00}$ and $N_{11}$ the numbers of pairs of observations that are in different clusters and in the same cluster for both, respectively. 
Denote by $N_{01}$ and $N_{10}$ the numbers of pairs that are in the same cluster for the truth but different clusters for the employed method, and in different clusters for the truth but the same cluster for the employed method, respectively.
The ARI, which is $0$ for random clustering and $1$ if both clusterings correspond, is then defined as
$$\text{ARI}=\frac{2(N_{00}N_{11}-N_{01}N_{10})}{(N_{00}+N_{01})(N_{01}+N_{11})+(N_{00}+N_{10})(N_{10}+N_{11})}.$$

The ARI measures the information shared by both clusterings. Denote the number of data points in cluster $i$ for the ground truth labels and in cluster $j$ for the clustering algorithm as $n_{ij}$, and define
\begin{eqnarray*}
M_{0}&=&-\sum_i\frac{\sum_j n_{ij}}{n}\log\left(\frac{\sum_j n_{ij}}{n}\right)\\
M_{1}&=&-\sum_j\frac{\sum_i n_{ij}}{n}\log\left(\frac{\sum_i n_{ij}}{n}\right)\\
M_{01}&=&\sum_i\sum_j\frac{n_{ij}}{n}\log\left(\frac{n_{ij}n}{(\sum_j n_{ij})(\sum_i n_{ij})}\right).
\end{eqnarray*}

The AMI is then given as $$\text{AMI}=\frac{M_{01}-\mathbb{E}[M_{01}]}{\text{Max}(M_0,M_1)-\mathbb{E}[M_{01}]},$$
where the expectation is taken over a random clustering under a hypergeometric model. The AMI is bounded by $(0,1)$, similar to the ARI.
The ARI and AMI are  reasonable when comparing partitions with different numbers of clusters, while we use the MR only when both partitions have the same number of clusters.

\subsection{Results and model architecture} \label{supp: results}
Selecting the architecture is crucial when working with DMFAs.  For the VIdmfa approach, 
the architecture was selected as described in the original paper testing all models 
with factor dimensions fulfilling the Anderson-Rubin condition with $L=2$ or $L=3$ layers. 
Each deep layer was initialized with $K^{(l)}=5$ components. For Scenario S1, for every replicate 
a model with $L=2$ layers and only one component in the deeper layer was selected, 
which is expected, because the clusters are Gaussian. For the cyclic data, a model with $L=2$ layers was selected 
for every replicate, but the dimension for the deeper layer differed depending on the particular replicate. 
For all real datasets, a model with $L=2$ layers was selected. For the \textbf{Olive data} and the \textbf{Ecoli data}, the only model fulfilling the Anderson-Rubin conditions has dimensions $D^{(1)}=3, D^{(2)}=1$. This model was also selected for the \textbf{Wine data}. 
For the \textbf{Vehicle data} a model with dimensions $D^{(1)}=4, D^{(2)}=1$ was selected and for the $\textbf{Satellite data}$ dimensions $D^{(1)}=11, D^{(2)}=2$ were selected. Only for the \textbf{Wine data} and the \textbf{Satellite data} was a model with more then one component in the deeper layer selected, having $K^{(2)}=2$ components for the \textbf{Wine data} and $K^{(2)}=4$ components for the \textbf{Satellite data}. These models are quite different to those selected in \cite{viroli2019deep} for the EMdgmm approach. For the latter, the architecture was selected as described in \citet{viroli2019deep} by fitting all possible models with $L=2$ or $L=3$ layers and a varying number of components $K^{(l)}=1,\dots5$ and selecting the best fit according to the BIC. 
 Due to its prior choices the VIgmm method  the optimal number of clusters can be found in a similar fashion as VIdmfa. Finally, for FMA and MFA the dimension of the latent variables was selected by the BIC.
Figure~\ref{fig: boxplot_S1_full} and Figure~\ref{fig: boxplot_S2_full} show the results on the two scenarios over all benchmark methods.

\begin{figure}
\center
\includegraphics[width=0.95\columnwidth,keepaspectratio]{images/simulated data/boxplot_S1_full.png}
\caption{\small Scenario S1. Boxplots summarize row-wise the MR, ARI and AMI across $100$ replications.}
\label{fig: boxplot_S1_full}
\end{figure}

\begin{figure}
\center
\includegraphics[width=0.95\columnwidth,keepaspectratio]{images/simulated data/boxplot_S2_full.png}
\caption{\small Scenario S2. Boxplots summarize row-wise the MR, ARI and AMI across $100$ replications. Note that not all benchmark methods converge on this data set.}
\label{fig: boxplot_S2_full}
\end{figure}

\FloatBarrier
\newpage
\singlespacing
\bibliography{litliste}
%\newpage
%\input{tabs}
%\newpage
%\input{figs}